\documentclass[acmsmall]{acmart}
\newif\ifshaphered
\ifdefined\isshaphered
\shapheredtrue
\fi

\usepackage{amsmath}
\usepackage{amsthm}
\usepackage{cases}
\usepackage{utfsym} 
\usepackage{bbding}
\newtheorem{theorem}{Theorem}
\newtheorem{definition}{Definition}
\newtheorem{lemma}{Lemma}
\usepackage[ruled]{algorithm2e}
\usepackage{booktabs}
\usepackage{multirow}
\usepackage{caption}
\usepackage{enumitem}
\usepackage{makecell}
\usepackage{hyperref}
\usepackage{mdframed}
\usepackage{xcolor}
\usepackage{subcaption}
\usepackage{tabularx}
\usepackage{enumitem}
\usepackage{diagbox}
\usepackage{array}
\usepackage{float}
\usepackage{soul}
\usepackage{comment}
\usepackage[normalem]{ulem}
\usepackage{fontawesome5}
\usepackage[table]{xcolor}
\setlist[itemize]{leftmargin=*}
\usepackage[noabbrev,capitalize]{cleveref}

\sethlcolor{yellow}

\SetKwComment{CommentTri}{$\triangleright$~}{}
\def \privsyn{\text{PrivSyn}\xspace}
\def \rap{\text{RAP}\xspace} 
\def \rapp{\text{RAP++}\xspace} 
\def \privmrf{\text{PrivMRF}\xspace}
\def \aim{\text{AIM}\xspace}
\def \gsd{\text{Private-GSD}\xspace}
\def \gem{\text{GEM}\xspace}
\def \merf{\text{DP-MERF}\xspace}
\def \ddpm{\text{TabDDPM}\xspace}
\def \dkl{\mathbb{D}_{\text{KL}}}

\newcommand{\mtdtt}[1]{\text{#1}}

\newmdenv[
  backgroundcolor=gray!10,  % 背景颜色
  linecolor=black,          % 边框颜色
  linewidth=1pt,            % 边框宽度
  innertopmargin=8pt,       % 内容上方内边距
  innerbottommargin=8pt,    % 内容下方内边距
  innerleftmargin=10pt,     % 内容左侧内边距
  innerrightmargin=10pt,    % 内容右侧内边距
  frametitlefont=\bfseries, % 标题字体加粗
  frametitlebackgroundcolor=gray!10, % 标题背景颜色与正文一致
  frametitlebelowskip=-1pt
]{keytakeaway}

\ifshaphered
\usepackage{longtable}
\newcommand\Rone[1]{\textcolor{orange}{#1}}
\newcommand\Rtwo[1]{\textcolor{blue}{#1}}
\newcommand\Rthree[1]{\textcolor{purple}{#1}}
\else
\newcommand\Rone[1]{#1}
\newcommand\Rtwo[1]{#1}
\newcommand\Rthree[1]{#1}
\fi
\AtBeginDocument{%
  }

\setcopyright{cc}
\setcctype{by}
\acmJournal{PACMMOD}
\acmYear{2025} \acmVolume{3} \acmNumber{6 (SIGMOD)} \acmArticle{299} \acmMonth{12} \acmPrice{}\acmDOI{10.1145/3769764}

\begin{document}

%%
%% The "title" command has an optional parameter,
%% allowing the author to define a "short title" to be used in page headers.
\title{Benchmarking Differentially Private Tabular Data Synthesis: [Experiments \& Analysis]}

%%
%% The "author" command and its associated commands are used to define
%% the authors and their affiliations.
%% Of note is the shared affiliation of the first two authors, and the
%% "authornote" and "authornotemark" commands
%% used to denote shared contribution to the research.
\author{Kai Chen}
\affiliation{
  \institution{University of Virginia}
  % \city{Charlottesville}
  % \state{Virginia}
  \country{USA}
  }
\email{kaichen@virginia.edu}

\author{Xiaochen Li}
\affiliation{
  \institution{UNC Greensboro}
  % \city{Greensboro}
  % \state{North Carolina}
  \country{USA}
  }
\email{X_LI12@uncg.edu}

\author{Chen Gong}
\affiliation{
  \institution{University of Virginia}
  % \city{Charlottesville}
  % \state{Virginia}
  \country{USA}
  }
\email{chengong@virginia.edu}

\author{Ryan McKenna}
\affiliation{
  \institution{Google Research}
  % \city{Seattle}
  % \state{Washington}
  \country{USA}
  }
\email{mckennar@google.com}
 
\author{Tianhao Wang}
\affiliation{
  \institution{University of Virginia}
  % \city{Charlottesville}
  % \state{Virginia}
  \country{USA}
  }
\email{tianhao@virginia.edu}

%%
%% By default, the full list of authors will be used in the page
%% headers. Often, this list is too long, and will overlap
%% other information printed in the page headers. This command allows
%% the author to define a more concise list
%% of authors' names for this purpose.
\renewcommand{\shortauthors}{Kai Chen, Xiaochen Li, Chen Gong, Ryan McKenna, and Tianhao Wang}

%%
%% The abstract is a short summary of the work to be presented in the
%% article.
\begin{abstract}
Differentially private (DP) tabular data synthesis generates artificial data that preserves the statistical properties of private data while safeguarding individual privacy. The emergence of diverse algorithms in recent years has introduced challenges in practical applications, such as inconsistent data processing methods, the lack of in-depth algorithm analysis, and incomplete comparisons due to overlapping development timelines. These factors create significant obstacles to selecting appropriate algorithms. 

In this paper, we address these challenges by proposing a benchmark for evaluating tabular data synthesis methods. We present a unified evaluation framework that integrates data preprocessing, feature selection, and synthesis modules, facilitating fair and comprehensive comparisons. Our evaluation reveals that a significant utility-efficiency trade-off exists among current state-of-the-art methods. Some statistical methods are superior in synthesis utility, but their efficiency is not as good as most deep learning-based methods. Furthermore, we conduct an in-depth analysis of each module with experimental validation, offering theoretical insights into the strengths and limitations of different strategies. Our code is open-sourced via the link.\footnote{\url{https://github.com/KaiChen9909/tab_bench}}
\end{abstract}

%%
%% The code below is generated by the tool at http://dl.acm.org/ccs.cfm.
%% Please copy and paste the code instead of the example below.
%%


%%
%% Keywords. The author(s) should pick words that accurately describe
%% the work being presented. Separate the keywords with commas.

% Articles V3mod284-V3mod334 use
\received{April 2025}
\received[revised]{July 2025}
\received[accepted]{August 2025}

%%
%% This command processes the author and affiliation and title
%% information and builds the first part of the formatted document.
\maketitle

\section{Introduction}

Private tabular data synthesis generates artificial data that preserves the statistical properties of real data while protecting individual privacy. 
This critical problem has a broad range of applications in practice, extending from healthcare~\cite{sangeetha2022differentially, khokhar2023differentially} to governmental planning~\cite{barrientos2018providingaccessconfidentialresearch, Cunningham_2021} and beyond.
For instance, in healthcare, there is a need to share patient data for medical treatment while preserving privacy. 
Similarly, in government, data analysts must analyze sensitive personal attributes while ensuring confidentiality.
Addressing this challenge has thus received much attention. 

Differential privacy (DP) has become the gold standard for protecting privacy. DP ensures that the inclusion or exclusion of any single data point does not significantly affect the outcome, thereby protecting each individual data point within a dataset. 
% The application of DP mechanisms to tabular data generation algorithms has received widespread research attention in the past. 
A substantial body of research has been proposed to address the data synthesis problem with DP. Based on their working principles, they can be broadly classified into two categories: statistical methods and deep learning methods. Statistical methods~\cite{zhang2017privbayes,vietri2022private,zhang2021privsyn,liu2023generating,zhang2017privbayes, mckenna2021winning, vietri2020neworacleefficientalgorithmsprivate, hardt2012simple} compress data information through statistical properties, such as low-dimensional data distributions, to achieve data generation. On the other hand, deep learning methods leverage deep learning frameworks designed for data generation, such as generative adversarial networks~\cite{liu2021iterative, harder2021dp,jordon2018pate,xie2018differentially, torkzadehmahani2019dp,gong2025dpimagebench} and diffusion models~\cite{kotelnikov2023tabddpm, pang2024clavaddpm,li2024privimage}.

In addition to these efforts to address this problem, many works also focus on providing comprehensive benchmarks. For instance, Du et al.~\cite{du2024towards} and Tao et al.~\cite{tao2022benchmarkingdifferentiallyprivatesynthetic} try to evaluate current methods under the same setting. But their works do not involve recent algorithms and thus lack completeness. Some other benchmark works~\cite{hu2023sokprivacypreservingdatasynthesis, yang2024tabular, fan2020survey} focus more on making algorithm analysis and comparison, lacking necessary empirical validation. 

In summary, even though several studies have been conducted in this field, we still face several challenges:
(1) \emph{Lack of unified evaluation settings.} Beyond algorithmic strategies, evaluation settings, such as data preprocessing, play a significant role in determining algorithm performance. However, these settings are often considered trivial and thus are frequently overlooked in many methods, which potentially leads to unfair comparisons between methods. 
(2) \emph{Lack of systematic and in-depth analysis.}
Current works often focus on proposing new methods, only providing limited intuition about algorithm analysis, or they only provide an overall comparison without detailed analysis, such as comparing deep learning and statistical methods' utility, but not delving into the reasons for the observed differences. Consequently, certain aspects of in-depth analysis, such as analysis for individual algorithm modules, remain underexplored. 
For example, questions such as how to select marginals more accurately are insufficiently addressed. However, many generative algorithms rely entirely on the selection of marginals, making this result particularly important.
(3) \emph{Lack of comprehensive comparison.}
Due to various reasons, such as concurrent development or relatively recent introduction, comparisons between recently proposed methods and existing works remain incomplete, particularly between some representative methods like \gsd~\cite{liu2023generating} and \aim~\cite{mckenna2022aim}. Furthermore, current comparisons largely focus on the overall utility of algorithms, with little attention given to experiments analyzing specific working modules.

In light of the above challenges, we believe that proposing a new benchmark for evaluating tabular data synthesis is necessary. The contributions of our benchmark work are as follows.

\vspace{0.5mm}
\noindent\textbf{Proposing a Unified Framework for Evaluation.}
We first propose a generalized framework and align all methods within this framework to ensure fair and objective comparisons. 
The framework consists of a {\it data preprocessing module}, a {\it feature selection module}, and a {\it data synthesis module}. Notably, this is the first framework to explicitly consider the impact of preprocessing on algorithm comparisons. Moreover, we move forward on the selection and synthesis modules by categorizing them according to their working principle, providing a new perspective to understand them better.
% the division of feature selection and data synthesis modules facilitates the analysis of their respective impacts on the overall algorithm performance.

\vspace{0.5mm}
\noindent\textbf{Providing Rigorous Analysis for Current Methods.}
Given our unified framework, we conduct an in-depth analysis of different modules. We consider different preprocessing methods and their drawbacks and advantages. For the feature selection module, we discuss the importance of introducing a scale penalty term during selection and prove the superiority of the adaptive marginal selection strategy. Finally, we investigate the efficiencies of current synthesis methods and discuss their potential limitations.

\vspace{0.5mm}
\noindent\textbf{Conducting a Comprehensive Comparison.}
We include current state-of-the-art methods~\cite{mckenna2022aim, zhang2021privsyn, cai2021data, vietri2022private, liu2021iterative, harder2021dp} under both statistical and deep learning methods, and newly proposed methods~\cite{liu2023generating, kotelnikov2023tabddpm} that have not been thoroughly explored.
Moreover, our evaluation is more fine-grained and helps us understand how each module of the algorithms functions independently.
% \xl{This is typically emphasized in the last sentence of the abstract and should not be placed here.}

We observed some important experimental findings, i.e., (1) A significant trade-off between utility and efficiency exists among current methods \Rtwo{under current implementations}. Two statistical methods, \aim~\cite{mckenna2022aim} and \privmrf~\cite{cai2021data}, outperform in utility but show worse time efficiency. Deep learning methods, even though relatively inferior in utility, are highly efficient. (2) Preprocessing is crucial for improving algorithm efficiency without significant synthesis errors. It is also algorithm-dependent, with different techniques better suited to specific synthesis methods. (3) All current synthesis modules exhibit limitations in different ways, such as low efficiency or inferior utility.

\section{Problem Formulation}
\label{sec: prelim}

Differential privacy (DP) has become the \textit{de facto} standard for data privacy.  It allows aggregated statistical information to be extracted while limiting the disclosure of information about individuals. More formally, the definition of DP is given by:

\begin{definition}[Differential Privacy] 
\label{def:dp}
An algorithm $\mathcal{A}$ satisfies $(\varepsilon,\delta)$-differential privacy ($(\varepsilon,\delta)$-DP) if and only if for any two neighboring datasets $D$ and $D'$ and any $T\subseteq \text{Range}(\mathbf{A})$, we have
$$
\Pr{\mathcal{A}(D) \in T} \leq e^{\varepsilon}\, \Pr{\mathcal{A}(D') \in T} + \delta. 
$$
\end{definition}
\noindent Here, we say two datasets are neighboring ($D \simeq D'$) when they differ on one tuple/sample. To achieve DP in different scenarios, many mechanisms have been employed, such as Gaussian mechanism~\cite{mironov2017renyi}, exponential mechanism~\cite{mckenna2022aim, liu2021iterative}, and DP-SGD~\cite{mironov2019r}. We provide a detailed introduction to them in the appendix of our full paper~\cite{chen2025benchmarking}.
In our work, we use R\'enyi DP as a tight composition tool, defined as:

\begin{definition}[R\'enyi DP~\cite{mironov2017renyi}] 
\label{def:rdp}
We say that an algorithm $\mathcal{A}$ satisfies $(\alpha, \varepsilon)$-R\'enyi DP ($(\alpha, \varepsilon)$-RDP) if and only if for any two neighboring datasets $D$ and $D'$ 
\[
D_{\alpha}(\mathcal{A}(D)||\mathcal{A}(D')) \leq \varepsilon,
\]
where $D_{\alpha}(Y||N) = \frac{1}{\alpha-1} \ln{\mathbb{E}_{x \sim N} \left[\frac{Y(x)}{N(x)}\right]^{\alpha}}$.
\end{definition}

RDP has composition and post-processing properties~\cite{mironov2017renyi}, which make it a suitable choice for complex algorithm design. Moreover, an $(\alpha, \varepsilon)$-RDP guarantee can easily be converted to a $(\varepsilon', \delta)$-DP guarantee via \cref{thm:rdp2dp} \cite{mironov2017renyi}. 
\begin{theorem} \label{thm:rdp2dp}
    If $f$ is an $(\alpha, \varepsilon)$-RDP mechanism, then it also satisfy $\left(\varepsilon + \frac{\log{1/\delta}}{\alpha - 1}, \delta\right)$-DP for any $0 < \delta < 1$.
\end{theorem}

One promising application of DP is for tabular data synthesis, wherein an artificial tabular dataset is generated that mirrors the statistical characteristics of the original dataset without compromising individual privacy. More formally, assuming that we have a dataset $D$ composed of $n$ records $\{x_1, \cdots, x_n\}$, and each record has $d$ attributes $\{A_1, \cdots, A_d\}$, we want to generate a dataset $D_s$ similar to $D$. The two datasets are considered similar based on some similarity metric, such as the $\ell_1$ distance or the performance under a downstream task (e.g., answering a range query or training a classification task). We give more concrete metrics in \cref{sec: exp setup}.

\section{Existing Work}
\label{sec: exist work}

This section examines existing DP tabular synthesis methods and identifies shortcomings in current benchmarks. These limitations inspire the development of our new benchmark.

\subsection{Existing Algorithms}

Broadly, existing DP tabular data synthesis methods can be divided into two categories, which are statistical methods and deep learning methods. 

\vspace{1mm}
\noindent \textbf{Statistical Methods}. The exploration of statistical methods started earlier. Shortly after the development of DP, researchers began investigating the data generation problem under the DP. \mtdtt{MWEM}~\cite{hardt2012simple} and \mtdtt{DualQuery}~\cite{gaboardi2015dualquerypracticalprivate} both release data by repeatedly improving an approximated distribution using Multiplicative Weight approach~\cite{hardt2010multiplicative}. Another notable line of research involves Bayesian networks, such as \mtdtt{PrivBayes}~\cite{zhang2017privbayes} and \mtdtt{BSG}~\cite{bindschaedler2017plausibledeniabilityprivacypreservingdata}. In addition, Li et al.~\cite{li2014differentially} try to utilize Copula functions for data generation.

In 2018 and 2020, NIST~\cite{url:nistchallenge2018, url:nistchallenge2020} hosted challenges about DP data synthesis. Among the competing algorithms, \mtdtt{PrivBayes}, \mtdtt{MST}~\cite{mckenna2021winning}, and \mtdtt{DPSyn}~\cite{li2021dpsynexperiencesnistdifferential} exhibited the best performance. All of these methods attempt to privately identify and answer highly correlated low-dimensional marginals. Their main differences lie in their methodology for selecting these low-dimensional marginals and in how they represent the data distribution from these noisy marginals. For example, \mtdtt{MST} synthesizes data using probabilistic graphical models (PGMs)~\cite{mckenna2019graphical}, \mtdtt{PrivBayes} uses a Bayesian model, and \privsyn iteratively updates an initialized dataset using an algorithm they call GUM.

After these NIST challenges, more advanced statistical methods were proposed. Zhang et al. proposed \privsyn~\cite{zhang2021privsyn} by organizing and refining \mtdtt{DPSyn}. Cai et al. introduced \privmrf~\cite{cai2021data}, and McKenna et al. proposed \aim~\cite{mckenna2022aim}, both of which dynamically select low-dimensional marginals and employ PGMs for synthesis. Moreover, methods such as \mtdtt{FEM}~\cite{vietri2020neworacleefficientalgorithmsprivate} and \rap/\rapp~\cite{aydore2021differentially,vietri2022private} treat synthesis as an optimization problem, utilizing FTPL~\cite{kalai2005efficient,suggala2020online,syrgkanis2016efficient} and relaxed projection~\cite{neel2018useheuristicsdifferentialprivacy}, respectively, to refine initialized datasets using adaptively selected marginals. More recently, Liu et al.~\cite{liu2023generating} proposed \gsd, which can apply genetic algorithms to adjust datasets based on any selected marginals iteratively.

\vspace{1mm}
\noindent \textbf{Deep learning Methods}. In addition to statistical methods, deep learning models have been widely explored for DP tabular data synthesis. In this work, we use the category ``deep learning model" to broadly encompass all neural network-related approaches, even though some of them do not have a ``deep" network structure. In NIST 2018, there was an effort to utilize GANs to generate data. However, this method, called DP-GAN~\cite{xie2018differentially}, did not demonstrate good performance. Further attempts on generative adversarial networks (GANs), such as DP-GAN~\cite{xie2018differentially}, \mtdtt{DP-WGAN}~\cite{srivastava2019differentially}, DP-CGAN~\cite{torkzadehmahani2019dp}, PATE-GAN~\cite{jordon2018pate}, and DP-CTGAN \cite{fang2022dp}, also demonstrated limited performance before. Therefore, we omit the comparison of this class of methods.

{
\setlength{\abovecaptionskip}{3pt}
\setlength{\belowcaptionskip}{-1pt}
\begin{table*}[t]
\footnotesize
    \centering
    \label{method}
    \caption{Summary of benchmarked algorithms. We match different algorithms within our framework and summarize/categorize their working pipelines. Here, `unspecified' means that related information is not mentioned or does not have a deterministic setting in the original paper. The original baselines column lists some other well-received baseline algorithms.
    % \tw{do you want to say we augment the 'select-noise-synthesis' pipeline into 4 steps?}
    }
    \vspace{-1mm}
    \resizebox{\textwidth}{!}{
    \begin{tabular}{c|c|c|c|c|c}
    \toprule 
        \textbf{Category} & \textbf{Method} & \textbf{Data Preprocessing} & \textbf{Feature Selection} & \textbf{Data Synthesis} & \textbf{Original Baselines} \\ \midrule
         % & MST~\cite{mckenna2021winning} & unspecified & non-adaptive & {  model-adjusting  } & PrivBayes, DPSyn \\
        \multirow{6}{*}{\makecell{Statistical\\ Method}} & \rap~\cite{aydore2021differentially} & Unspecified & Adaptive & Relaxed Projection & \mtdtt{FEM}~\cite{vietri2020neworacleefficientalgorithmsprivate},\mtdtt{HDMM}~\cite{McKenna_2018} \\
        & \privsyn~\cite{zhang2021privsyn} & Categorical Preprocessing & Non-adaptive & GUM & \mtdtt{PrivBayes}~\cite{zhang2017privbayes}, \mtdtt{DualQuery}~\cite{gaboardi2015dualquerypracticalprivate},\mtdtt{PGM}~\cite{mckenna2019graphical}\\ 
        & \privmrf~\cite{cai2021data} & Unspecified & Adaptive & PGM & \mtdtt{PrivBayes}, \mtdtt{BSG}~\cite{bindschaedler2017plausibledeniabilityprivacypreservingdata}, 
        \mtdtt{DP-WGAN}~\cite{srivastava2019differentially}, \mtdtt{DP-Copula}~\cite{li2014differentially} \\ %, \mtdtt{PGM}\\ 
        & \rapp~\cite{vietri2022private} & Unspecified & Adaptive & Relaxed Projection & \rap, \merf, DP-CTGAN~\cite{fang2022dp}, \mtdtt{PGM}\\
        & \aim~\cite{mckenna2022aim} & Unspecified & Adaptive & PGM & \privmrf, \rap, \mtdtt{MST~\cite{mckenna2021winning}}, \mtdtt{MWEM}~\cite{hardt2012simple}\\
        & \gsd~\cite{liu2023generating} & Unspecified & Unspecified & Genetic Algorithm & \gem, \rapp, \mtdtt{PGM}~\cite{mckenna2019graphical}\\

        \midrule
        \multirow{3}{*}{\makecell{deep learning\\ Method}} & \gem~\cite{liu2021iterative} & Unspecified & Adaptive & Generative Network & \rap, \mtdtt{MWEM}, DualQuery\\
        & \merf~\cite{harder2021dp} & Unspecified & Non-adaptive & Generative Network & \mtdtt{DP-GAN}~\cite{xie2018differentially}, \mtdtt{DP-CGAN}~\cite{torkzadehmahani2019dp} \\ 
        & \ddpm~\cite{kotelnikov2023tabddpm} & Unspecified & - & Diffusion Model & -\\
        % & DP-LLMGen~\cite{tran2024differentially} & - & - & model-adjusting & \rap, \rapp, \gsd\\
    \bottomrule
    \end{tabular}
    }
    \vspace{-1.5mm}
    \label{algo summary}
\end{table*}
}

More recent deep learning approaches, including \gem~\cite{liu2021iterative} and \merf~\cite{harder2021dp}, represent generative network-based advancements. \gem combines generative networks with adaptive marginal selection mechanisms, while \merf employs random Fourier feature loss to train generative networks. Besides, \ddpm~\cite{kotelnikov2023tabddpm} leverages diffusion models' representational power to fit target data directly. While \ddpm was not originally designed for DP, it achieves state-of-the-art performance among non-DP methods and can be adapted for DP synthesis using DP-SGD~\cite{du2024towards}. Consequently, we include \ddpm in our analysis.

Our work focuses on recently proposed methods that have not been well compared previously and those representing the current state-of-the-art, as summarized in~\cref{algo summary}. Here, we must emphasize that specifically, we don't consider some LLM-based methods~\cite{tran2024differentially,sablayrolles2023privatelygeneratingtabulardata}. That is because LLMs are trained on extensive public datasets, which will introduce evaluation bias and affect our further comparison of algorithm modules.

\subsection{Existing Benchmark Works}

In addition to the proposed algorithms, there are several benchmark studies~\cite{du2024towards, hu2023sokprivacypreservingdatasynthesis, yang2024tabular, fan2020survey, tao2022benchmarkingdifferentiallyprivatesynthetic} that investigate the problem of DP data synthesis. However, previous works all have some weaknesses, summarized as follows.
\begin{itemize}[leftmargin=*, label = \textbullet] 
\item \textbf{Lack of unified comparison setting}. Du et al. ~\cite{du2024towards} have made an experimental evaluation of current works, but ignore the importance of a unified setting (e.g., preprocessing), which may significantly influence the comparison fairness. 
\item \textbf{Not include recent advanced works}. Du et al. ~\cite{du2024towards} and Tao et al.~\cite{tao2022benchmarkingdifferentiallyprivatesynthetic} focus on utility evaluation, but they both lack investigation for some advanced methods, such as \rapp and \gsd.
Fan et al.'s work ~\cite{fan2020survey} only focuses on GAN-based methods. 
\item \textbf{Lack of deep analysis}. Yang et al.~\cite{yang2024tabular} provide a survey of many methods and further investigate distributed data synthesis. However, they do not delve deep into these algorithms' working principles. Du et al.'s work and Tao et al.'s work also have a similar weakness in lacking deep algorithm analysis. 
\item \textbf{Lack of comprehensive experiments}. Hu et al.~\cite{hu2023sokprivacypreservingdatasynthesis} provide analysis for a wide range of DP synthesis algorithms, not only about tabular data synthesis but also trajectory data and graph data. This work, though trying to make an in-depth analysis, lacks comprehensive experiments. 
\end{itemize}

Except for these works, there are other ``benchmark-like" works on different aspects of this problem. For instance, Ganev et al.~\cite{ganev2025importancediscretemeasuringimpact} discuss the importance of discretization in tabular data synthesis. Moreover, Stadler et al.~\cite{stadler2022syntheticdataanonymisation} focus more on the quantitative evaluation of privacy gain. These works provide in-depth research from different perspectives but lack a more comprehensive viewpoint. 

{
\setlength{\abovecaptionskip}{3pt}
\setlength{\belowcaptionskip}{-1pt}
\begin{figure}[t]
    \centering
    \includegraphics[width=0.55\linewidth]{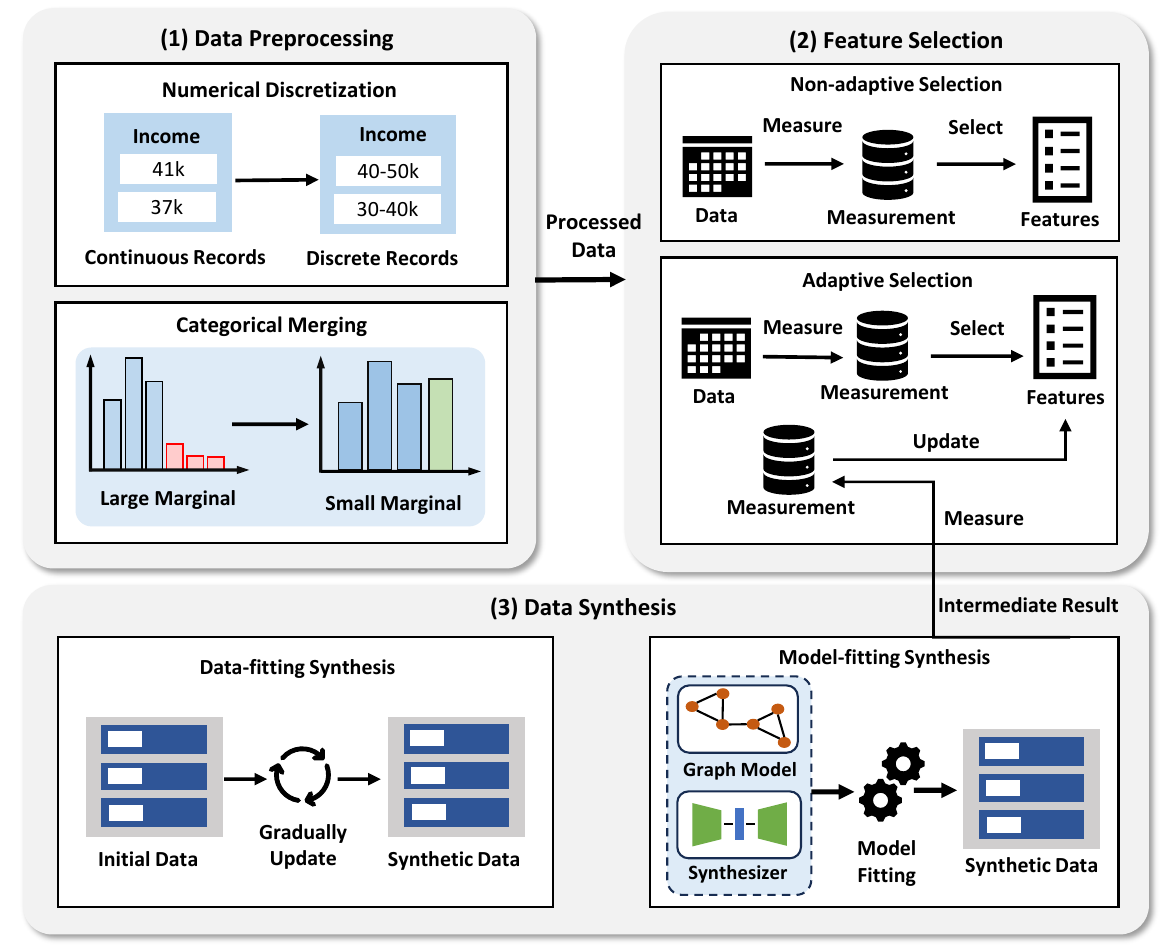}
    \caption{The proposed unified framework. The dataset is first preprocessed and then represented by some selected features.
    Finally, using the selected features, the synthesis algorithm generates data as the output of the workflow.}
    \vspace{-1mm}
    \label{fig:framework}
\end{figure}
}

Realizing the weaknesses of current research, our work aims to address these issues by proposing a standardized algorithmic framework (\cref{sec: framework}), providing rigorous analysis (\cref{sec: prep}, \cref{sec: select}, and \cref{sec: syn}), and conducting detailed experiments (\cref{sec: exp}).

\section{Framework Overview}
\label{sec: framework}

Some previous works~\cite{liu2021iterative,mckenna2021winning,hu2023sokprivacypreservingdatasynthesis} have identified that current methods have several common working patterns and proposed unifying algorithmic frameworks, focusing primarily on feature selection and dataset synthesis. 
% \gc{[Chen: please highlight the importance of pre-processing here, give readers some intuition why we should incorporate this module.]} 
However, when dealing with a complex dataset, preprocessing it into a more manageable form should also be a necessary part of the algorithm.
As presented~\cref{fig:framework}, our framework extends these approaches by incorporating a data preprocessing module. This section gives an overview of our framework and presents how modules work together to form a complete and cohesive data synthesis pipeline.

\vspace{0.5mm}
\noindent \textbf{Preprocessing.} As observed in our evaluation, many datasets contain attributes with large domain sizes (e.g., exceeding $10^5$ in Loan dataset~\cite{loandata}), which pose significant challenges for data synthesis. 
For statistical methods, large attribute domains lead to expansive, low-dimensional marginals, which will introduce excessive DP noise during synthesis and degrade performance. 
% However, most prior works~\cite{mckenna2022aim, liu2021iterative, vietri2022private, liu2023generating, cai2021data} fail to deal with such cases.
For deep learning methods, a large domain requires a synthesizer with a substantial model size to learn the relationship between different attributes, leading to slower training and higher resource demands. Additionally, larger model sizes require a higher scale of DP noise under the same privacy budget compared to smaller synthesizers, degrading synthetic performance~\cite{gong2025dpimagebenchunifiedbenchmarkdifferentially}.

% Ganev et al.~\cite{ganev2025importancediscretemeasuringimpact} proposed that these challenges can be alleviated by proper preprocessing. \gc{[Chen: List some figure here, like prepossessing improve xxx\% of xxx, compared to the method without considering it.]}
% Unfortunately, Ganev's work do not give us data without preprocessing (may be because large time complexity)
Therefore, preprocessing is important and non-negligible in algorithm workflows, which can reduce the dimensionality of marginals by effectively merging those with similar characteristics and compressing the domains of the attributes.
However, as shown in \cref{algo summary}, preprocessing is often overlooked by prior works~\cite{aydore2021differentially, mckenna2022aim, liu2021iterative, liu2023generating, kotelnikov2023tabddpm, harder2021dp, cai2021data}. This omission in the algorithm pipeline may affect the fairness of comparisons between methods.

\vspace{0.5mm}
\noindent \textbf{Feature Selection.}
Even after effective preprocessing, the domain of the whole dataset increases exponentially with the number of attributes, making methods that rely on a histogram representation computationally infeasible. A practical approach is to utilize some representative local data features, such as \emph{low-dimensional marginals}, to approximate the full joint data distribution~\cite{mckenna2022aim, zhang2021privsyn, cai2021data, liu2023generating}. Therefore, the second step in the framework is to measure the representativeness of the features and select necessary features. To further analyze how to better select features in \cref{sec: select}, we will categorize existing strategies into two categories:

\begin{enumerate}[label=\textbullet, leftmargin=*]
    \item \textbf{Non-adaptive Feature Selection}. A straightforward approach to this step involves performing one-shot feature selection. Some methods, such as \merf and \gsd, predefine a fixed set of features without selection. While others, like \privsyn, measure and select features in one step.
    \item \textbf{Adaptive Feature Selection}. Adaptive methods iteratively obtain new features based on intermediate feedback from previous selection steps to optimize algorithms. For example, \aim selects one new marginal at each iteration step and fits the model based on all selected marginals, which then serves as the reference for the next round of marginal selection. Moreover, DP-SGD iteratively calculates DP gradient features to update the model.
\end{enumerate}

% \vspace{0.5mm}

\noindent \textbf{Data Synthesis}. 
The third step in the framework is to synthesize data that aligns well with the features selected in the feature selection step. Current methods apply a wide range of algorithms to achieve this synthesis step. Broadly, there are two approaches: fitting the dataset or fitting the model. 

\privsyn manually adjusts data records to match the marginals, while \gsd achieves this by genetic algorithm~\cite{such2018deepneuroevolutiongeneticalgorithms}. \rap and \rapp use the relaxed projection mechanism to optimize the records of an initialized dataset. These methods are constructed by adjusting records. Some other methods fit models for data generation. PGM~\cite{mckenna2019graphical} is a classical graphical model for tabular data synthesis, which utilizes a tree-like model to represent the distribution of data. deep learning models such as generative networks and diffusion models have also been applied here.

It is also notable that this division is not mutually exclusive. We can regard an initialized dataset as a model where each data record is a model parameter, so adjusting data can also be regarded as fitting a model. Our characterization is primarily for explaining algorithm intuition.

\section{Preprocessing} 
\label{sec: prep}

As mentioned in \cref{sec: exist work}, when synthesizing complex datasets, we may encounter attributes with high cardinality. For example, an address or income attribute can have thousands of distinct values. Conducting statistics on such attributes is infeasible due to the high memory requirements and long execution times. Therefore, a preprocessing step is necessary for algorithm comparison. We surveyed some previous works~\cite{zhang2021privsyn, mckenna2021winning, ganev2025importancediscretemeasuringimpact} and summarized the data types requiring preprocessing into two types: categorical and numerical. Before introducing this section, we need to clarify a basic assumption: domain information is considered public knowledge. This assumption is reasonable in many cases. For example, the domains of personal attributes in census data are well documented on the IPUMS website~\cite{ipums}.

{
\setlength{\textfloatsep}{7pt}
\begin{algorithm}[!t]
\caption{DP Rare Category Merge}
\label{rare value merge}
\LinesNumbered
\KwIn{dataset $D$, merge threshold parameter $\theta$, unique value threshold $\beta$, DP parameter $\rho_2$}
\KwOut{preprocessed dataset $D$} 
$V_c \leftarrow $ categorical variables with domain size $\ge \beta$; \\
$\rho' \leftarrow \rho_2 / |V_c|$; \\
$\sigma = \sqrt{1/(2\rho')}$; \\
\For {$j \in V_c$}{
    $b \leftarrow $ 1-way marginal of attribute $A_j$; \\ 
    % $\sigma \leftarrow \sqrt{\frac{1}{2\rho'}}$ \\
    $\hat{b} = b + \mathcal{N}\left(0, \sigma^2\right)$; \\
    \For {$i = 1 : |\hat{b}|$} { 
        $\theta' \leftarrow \max \left\{\theta \cdot \sum\hat{b}, \; 3\sigma\right\}$ \CommentTri*[r]{Merging threshold}
        \If{$\hat{b}[i] < \theta'$}{
            replace $i$-th value with the rare encoding value;
        }
    }
}
\Return $D$
\end{algorithm}
}

\subsection{Categorical Attributes Preprocessing} 
Some prior works~\cite{zhang2021privsyn,mckenna2021winning} have proposed a $3\sigma$ merging strategy to preprocess categorical variables, where categories with counts below $3\sigma$ (where $\sigma$ represents the DP noise standard deviation in the counting process) are combined. 

By controlling each category's frequency to be large enough (larger than $3\sigma$), this approach helps mitigate the impact of noise. However, this method has limitations: when we have a large privacy budget, $3\sigma$ could be a small value. If we continue using $3\sigma$ as the threshold for combining, we may be able to achieve high accuracy, but we cannot reduce the attributes' domain size to ensure algorithm efficiency.

In response to these limitations, we improve the $3\sigma$ merging method, as shown in \cref{rare value merge}, by applying a dual merging threshold $\max\{3\sigma,\; \hat{n}\theta\}$.
Here $\hat{n}$ is the privately measured number of records in the dataset and $\theta$ is the threshold parameter. By introducing a fixed threshold, we can avoid the case when $\sigma$ is too small to reduce the attribute's domain complexity.

\subsection{Numerical Attributes Preprocessing}
\label{subsec: discretize}

Continuous numerical variables often exhibit dense distributions within specific intervals, with numerous unique values. Different from categorical attributes, these unique values have numerical correlation, making value combining (like what we do for categorical attributes) impossible. Thus, discretization is a proper preprocessing method for numerical attributes. Here, we outline two potential discretization approaches:

\vspace{0.5mm}
\noindent \textbf{Uniform Binning}.
Some previous works~\cite{Dick_2023, ganev2025importancediscretemeasuringimpact} use uniform binning for continuous data preprocessing. It partitions an attribute's domain into equal-length intervals, relying only on the attribute's domain range and a predefined number of bins. Formally, this method can be expressed as $\text{Uniform Bin}(x) = \left\lfloor \frac{x - x_{\ell}}{h} \right\rfloor$, where $x_{\ell}$ is the lower bound of the attribute's domain, and $h$ is the length of the uniform interval determined by the bin number. 

Uniform binning is advantageous because it requires no detailed data information, avoiding the need for additional privacy budget allocation. However, it has significant drawbacks, particularly for attributes with uneven distributions. For example, when data is highly concentrated around specific values, uniform binning can lead to inefficient binning, as it may allocate unnecessary bins to sparsely populated areas.

% \vspace{0.5mm}
% \noindent{\textbf{Exponential binning}}. For some data distributions, an exponential binning may be more natural.  Its relationship to uniform binning method is expressed as: 
% \[
% \text{Exponential Bin}(x) = \text{Uniform Bin}(\log_2{(x - x_{\ell})}).
% \]
% This approach smooths data distribution by compressing large ranges by logarithmic transformation while maintaining finer granularity in dense regions.

\begin{algorithm}[!t]
\caption{PrivTree Binning}
\label{PrivTree}
\LinesNumbered
\KwIn{dataset $D$, unique value threshold $\beta$, divide parameter $\theta$, privacy parameter $\rho_1$} 
\KwOut{preprocessed dataset $D$} 
Set $T = \emptyset$ \\
$V_n \leftarrow $ numerical variables with domain size $\ge \beta$\\
$\beta_0 \leftarrow 2$\\ 
$\lambda' \leftarrow \frac{2\beta_0 - 1}{\beta_0 - 1} \cdot \sqrt{\frac{|V_n|}{2\rho_1}}$ \\ 
$\delta' \leftarrow \lambda' \cdot \ln{\beta_0}$ \\
\For {$j \in V_n$}{
    $\mathcal{T} \leftarrow$ PrivTree($D[j], \lambda', \delta', \theta$) \\
    Append $\mathcal{T}$ to $T$
}
Apply $T$ to discretize dataset $D$\\
\Return $D$
\end{algorithm}

\vspace{0.5mm}
\noindent{\textbf{PrivTree}}. A limitation of uniform binning is its reliance on a predetermined number of bins, which introduces concerns about hyperparameter selection. To address this, PrivTree decomposition~\cite{zhang2016privtree, tao2022benchmarkingdifferentiallyprivatesynthetic} can be used as a self-adaptive discretization method.

PrivTree employs a tree structure to iteratively divide the domain of an attribute, with splits continuing until intervals contain only a small number of records. However, since PrivTree utilizes sensitive data for domain division, it requires a fraction of the privacy budget to guarantee differential privacy. We use PrivTree on multiple attributes whose domain size is larger than a threshold, with algorithm details in \cref{PrivTree}. We provide the proof of DP guarantee of this algorithm in the appendix of our full paper~\cite{chen2025benchmarking}.

\section{Feature Selection}
\label{sec: select}
% \tw{similar as preprocessing, feature selection is also important in stat methods but automated in ml methods. but of course, ml methods automated selection at a cost of a slower convergence rate (see, e.g., Differentially Private Learning Needs Better Features (or Much More Data))}

Selecting features determines the performance of many algorithms. Similar to preprocessing methods, it is important for statistical methods, while some deep learning methods can automatically learn the characteristics of the dataset by training models. In this section, we will briefly introduce the currently proposed selection methods and analyze them.

\subsection{Existing Selection Methods}
% We categorize the feature selection methods of existing works into two types: non-adaptive methods and adaptive methods, based on whether they iteratively select marginals according to some intermediate results from previous selection rounds.

\vspace{0.5mm}
\noindent \textbf{Non-adaptive Feature Selection}.
Non-adaptive methods perform all feature computations and operations at the beginning of the algorithm based on the characteristics of the marginals.
The selected features will then be delivered to the synthesis modules without any further refinement. 

Some algorithms directly predefine a set of features instead of paying attention to selecting representative features. \merf~\cite{harder2021dp} leverages random Fourier features to capture correlations among numerical variables, while employing 2-way marginals for categorical variables. In addition, some methods with strong fitting ability can work on all two-way marginals. For example, Liu et al.~\cite{liu2023generating} conduct experiments on all two-way marginals to demonstrate the performance of \gsd.

Instead of predefining some features, \privsyn~\cite{zhang2021privsyn} selects the most highly correlated marginals while respecting the privacy budget in one round. They measure each marginal using metric $\mathsf{InDif}_{i,j} = \left|M_{i,j}- M_{i} \times M_{j}\right|$,
where $M$ denotes the attribute marginal. The selection process involves minimizing the expected error
\[ 
    \displaystyle \sum_{i} \left( N_i x_i + \mathsf{InDif}_i (1-x_i)\right),
\]
where $x_i \in \{0, 1\}$ denotes the selection decision and $N_i$ is DP noise.

\vspace{0.5mm}
\noindent \textbf{Adaptive Feature Selection.}
Different from non-adaptive methods, adaptive methods continuously update feature selection with feedback from the previous selection steps. In each synthesis round, adaptive feature selection conducts many queries on the current estimation, and the features with large errors are selected. These features will be used for the selection of future rounds. This strategy is used by methods like \rap~\cite{aydore2021differentially}, \rapp~\cite{vietri2022private}, \gem~\cite{liu2021iterative}, \privmrf ~\cite{cai2021data} and \aim~\cite{mckenna2022aim}. These methods differ in several key aspects:
    \begin{itemize}[label=\textbullet, leftmargin=*] 
        \item First, in terms of initialization, \privmrf uses a carefully designed criterion to select a small set of features as the starting point, while \aim initializes with all 1-way marginals. In contrast, \rap, \rapp and \gem do not specify any initialization. 
        
        \item Secondly, the selection criteria also differ among these methods. Both \aim and \privmrf's marginal selection criteria include a punishment term proportional to the marginal scale, allowing for a balance between noise level and feature representativeness. Other methods measure the features without a penalty term, which is one-sided and potentially introduces more errors. 

        \item Finally, the selection mechanism varies: \aim and \gem utilize the exponential mechanism~\cite{cesar2021bounding}, whereas \rap and \rapp employ the Gumbel mechanism~\cite{aydore2021differentially}, allowing them to select more than one feature in each round; \privmrf takes a different approach, directly adding Gaussian noise to the selection criteria for selection and select the largest one.

    \end{itemize}

\subsection{Analysis for Selection Algorithms}
\label{subsec: adapt}

Some previous studies~\cite{mckenna2022aim, cai2021data, liu2021iterative} have discussed the important properties of a good selection mechanism. In this subsection, we formally investigate some aspects of it. Since all methods, except for \merf, which predefines features without selection, focus on marginal selection, we only discuss marginal selection here.

\vspace{1mm} 
\noindent \textbf{Importance of Scale Penalty Term}. For any marginal $M$, we have two choices: choose and privatize it or do not choose it. Let the marginal estimated by available information be $M_0$, and the privatized marginal be $\hat{M}$. We have $\hat{M} = M + \mathcal{N}(0, \sigma)$, where $\sigma$ is known privacy budget. The expected error can be derived as, 
\[
\left\{
\begin{aligned}
&\text{ Selection error: }\; \lVert M - \hat{M}\rVert_1 = n\sigma\sqrt{2/\pi} \\
&\text{ Unselection error: }\; \lVert M - M_0\rVert_1 .
\end{aligned}
\right.
\]
Therefore, we can compare these two errors, denoted as $\lVert M - M_0\rVert_1 - n\sigma\sqrt{\frac{2}{\pi}}$, to determine the selection result, which demonstrates the importance of the scale penalty term. This equation is also how some selection criteria involve the scale penalty term.

\vspace{1mm}
\noindent \textbf{Superiority of Adaptive Selection}. Establishing the superiority of the adaptive selection strategy is challenging in the absence of prior information about the generation module. Our analysis, therefore, proceeds under a set of reasonable assumptions. First, we consider the case of selecting 2-way marginals. Second, we assume that conditional independence is used for marginal selection. Specifically, assuming that before selecting $(A_i, A_j)$, we have already fitted marginals $(A_i, A_1, \cdots, A_k)$ and $(A_j, A_1, \cdots, A_k)$, so that we can use them to estimate the distribution of $(A_i, A_j)$ as 
\begin{equation} \label{eq: adapt measure}
\begin{aligned}
\hat{\Pr}[A_i, A_j] = \sum \Pr[&A_1, \cdots, A_k] \cdot \\ &\Pr[A_i | A_1, \cdots, A_k] \Pr[A_j | A_1, \cdots, A_k]. 
\end{aligned}
\end{equation}

Before selecting the marginal $(A_i, A_j)$, the non-adaptive methods do not have any intermediate results and can only use independent 1-way marginals to measure its representativeness. This independent measurement can be written as $\dkl\left(\Pr[A_i, A_j] \left\|\; \Pr[A_i]\Pr[A_j] \right.\right)$, where $\dkl$ is the KL Divergence~\cite{kullback1951information}. The adaptive methods, because they select features based on intermediate synthesis results, will have a conditional estimation as $\dkl\left(\Pr[A_i, A_j] \left\|\; \hat{\Pr}[A_i, A_j]\right.\right)$, which is the true KL divergence when choosing marginal $(A_i, A_j)$. Here $\hat{\Pr}[A_i, A_j]$ is defined in Equation~(\ref{eq: adapt measure}). Now we give the following theorem to prove the superiority of adaptive selection.
\begin{theorem} \label{theorem: adapt measure}
    For any pair of attributes $(A_i, A_j)$, the KL divergence error of conditional estimation is always no larger than that of independent estimation. Formally, we have
    \begin{equation*} \label{eq: theorem: adapt measure} 
    \begin{aligned}
    \dkl\left(\Pr[A_i, A_j] \left\|\; \hat{\Pr}[A_i, A_j]\right.\right) \leq \dkl\left(\Pr[A_i, A_j] \left\|\; \Pr[A_i]\Pr[A_j] \right.\right)
    \end{aligned}
    \end{equation*}
    Here $\hat{\Pr}[A_i, A_j]$ is defined in Equation~(\ref{eq: adapt measure}).
\end{theorem}

The proof of \cref{theorem: adapt measure} is deferred to the appendix of our full paper~\cite{chen2025benchmarking}. In essence, this theorem demonstrates that under our assumptions, non-adaptive methods tend to overestimate the representativeness of features under KL divergence, whereas adaptive methods can correct this error by leveraging intermediate results. The more general comparisons between adaptive and non-adaptive selection strategies will be conducted in our experiments. Another notable fact is that compared to the non-adaptive methods, the adaptive methods require multiple rounds of data synthesis or computation during the feature selection, which can lead to higher time consumption as a trade-off for their superiority in utility.

\section{Data Synthesis Module Comparison}
\label{sec: syn}

Previous sections have shown that current solutions utilize various techniques to generate data on selected features. This raises critical questions about their utility and efficiency. Thus, we will analyze this problem in this section. For data-fitting methods, there are \mtdtt{GUM}, \mtdtt{Genetic algorithm}, and \mtdtt{Relaxed} \mtdtt{Projection}. For the model-fitting type, we consider \mtdtt{PGM} and (deep) \mtdtt{generative} \mtdtt{network}.

% \vspace{0.5mm}
\vspace{0.5mm}
\noindent \textbf{\mtdtt{GUM}}. \mtdtt{GUM}, used by \privsyn, is an iterative adjustment method that modifies values in the initial dataset to align with the selected marginals. We assume that \mtdtt{GUM} merges marginals to $k$ cliques of sizes $\{c_1, \cdots, c_k\}$. Since the maximum number of operations to fit each value in the marginals is no more than the size of the synthetic dataset, the time complexity of \mtdtt{GUM} is $\mathcal{O}\left(\sum_{i=1}^k T c_i n\right)$, where $T$ is the number of update iterations, $n$ represents the synthetic dataset size, respectively. Because \mtdtt{GUM} strictly controls the clique size $c_i$, ensuring we have small cliques and simplifying the update process (e.g., we can choose a small $T$ to reach convergence). This guarantees the efficiency of \mtdtt{GUM}. However, it fits marginals one by one, overlooking overall correlations, limiting its utility.

% \vspace{0.5mm}
\vspace{0.5mm}
\noindent \textbf{\mtdtt{Genetic Algorithm}}. Genetic algorithms~\cite{such2018deepneuroevolutiongeneticalgorithms} use mutation and crossover operations to adjust datasets. The complexity depends on the number of mutations and crossovers per iteration. For instance, the algorithm by Liu et al.~\cite{liu2023generating} has a complexity of $\mathcal{O}\left(T(P_{m} + P_{c})\right)$, where $T$ is the number of iterations, and $P_{m}$ and $P_{c}$ represent mutations and crossovers per iteration, respectively. The execution time of genetic algorithms is strongly influenced by the number of tuning rounds, which often exceeds the dataset size when handling complex datasets with diverse values.

% \vspace{0.5mm}
\vspace{0.5mm}
\noindent \textbf{\mtdtt{Relaxed Projection}}. Relaxed projection mechanism~\cite{aydore2021differentially,vietri2022private} treats the dataset as a trainable model, optimizing it to match marginals. Given $T$ optimization rounds, synthetic data size $n$, and data dimension $d$, the time complexity is $\mathcal{O}(Tnd)$. 
This method can be regarded as both a model-fitting and data-adjusting approach. 
The complexity is driven by the size of the synthetic data and the number of optimization rounds. High-dimensional datasets with large attribute domains can result in an inflated encoded data dimension, making optimization difficult to converge and more time-intensive.

\vspace{0.5mm} 
\noindent \textbf{\mtdtt{PGM}}. \mtdtt{PGM}~\cite{mckenna2019graphical} constructs a junction tree of marginal cliques $C_1, C_2, \cdots, C_k$, ensuring that the intersection set $S_i$ of any two cliques appears only in those marginals. The overall distribution is approximated as:
\begin{equation} \label{eq:pgm} 
\Pr[A_1, \cdots, A_d] \approx \Pr[C_1] \cdot \prod_{i=2}^{k} \Pr[C_i \setminus S_i \; |\; S_i]. 
\end{equation}
Let $T$ be the number of training iterations, and let $k$ cliques have sizes $\{c_1, \cdots, c_k\}$. The total complexity is $\mathcal{O}\left(\sum_{i=1}^k Tc_i + nk\right)$, which includes model training and data genration complexity. \mtdtt{PGM}'s efficiency depends on constructing reasonable cliques, which are automatically determined by the junction tree. Densely selected marginals can lead to large cliques, greatly increasing time costs.

\vspace{0.5mm} 
\noindent \textbf{\mtdtt{Generative Network}}. Generative networks~\cite{liu2021iterative} rely on parameters $m$, batch size $b$, and training iterations $T$. The training complexity is $\mathcal{O}(Tmb)$. Generating $n$ synthetic records results in a total complexity of $\mathcal{O}(Tmb + mn)$. The efficiency is affected by network design and training hyperparameters, which can be challenging for high-dimensional datasets.

\vspace{0.5mm} 
In summary, we find that model-fitting type methods heavily depend on the construction of the generative model. While this approach can achieve high efficiency, it also risks encountering the curse of dimensionality. In contrast, data-fitting methods are often limited by the complexity of the data itself, which can significantly impact algorithm efficiency.

{
\setlength{\abovecaptionskip}{3pt}
\setlength{\belowcaptionskip}{0pt}
\begin{table}[t]
\footnotesize
    \centering
    \caption{\Rtwo{Summary of investigated datasets. 
    We report the number of records, the number of attributes, numerical attributes, categorical attributes, and minimum and maximum attribute domain size.
    }}
    \label{info:datasets}
    \resizebox{0.6\columnwidth}{!}{
    \begin{tabular}{l|ccccc}
        \toprule
        \textbf{Name} & \textbf{\#Records} & \textbf{\#Attr} & \textbf{\#Num} & \textbf{\#Cat} & \textbf{\makecell{Min/Max\\ Domain}} \\ 
        \midrule
        ACSincome (INC)~\cite{ding2021retiring} & 55320  & 10 & 2 & 8 & 2$\sim$93\\ 
        ACSemploy (EMP)~\cite{ding2021retiring} & 37881 & 17 & 1 & 16 & 2$\sim$92\\
        Bank (BK)~\cite{bank_marketing_222} & 45211& 16& 6& 10& 2$\sim$6024\\
        Higgs-small (HIG)~\cite{higgsdata} & 98049 & 28 & 28 & 1 & 2$\sim$73715\\
        Loan (LN)~\cite{loandata} & 134658 & 42 & 25 & 17 & 2$\sim$93995\\
        % Colorado & 661967 & 98 & 0 & 98 & \\ 
        \bottomrule
    \end{tabular}}
\end{table}
}

\section{Experimental Setup}
\label{sec: exp setup}

Our evaluations aim to (1) compare all well-established methods within our unified framework; (2) explore and verify the importance of preprocessing in DP tabular synthesis tasks; (3) investigate feature selection and synthesis modules of these methods for a more fine-grained comparison. The comparison results can validate previous theoretical analysis and guide method selection for different modules within the framework. \Rtwo{To achieve these goals, we design and conduct three main experiments. We summarize them and some important findings as follows.
\begin{enumerate}[leftmargin=*, label=\textbullet]
    \item \textit{Overall Evaluation}: We evaluate method performance across metrics and datasets. A key observation is that statistical methods like \aim and \privmrf demonstrate better synthesis utility, while under current implementations, deep learning methods can achieve higher time efficiency. 
    \item \textit{Preprocessing Investigation}: These experiments focus on the preprocessing investigation by comparing preprocessed datasets with raw ones. Through experiments, we prove that preprocessing is essential in reducing algorithm complexity without introducing much synthesis error. 
    \item \textit{Module Comparison}: Finally, we evaluate the effectiveness of different feature selection and synthesis modules by fixing one module and reconstructing algorithms by changing the other module. We find that (1) the adaptive selection strategy can help the algorithm achieve a higher synthesis utility; (2) existing different synthesis modules show their limitations in either utility or scalability.
\end{enumerate}
}

% To achieve this, we design three main experiments as follows, (1) \textit{Overall Evaluation}: We evaluate method performance across metrics and datasets; (2) \textit{Preprocessing Investigation}: These experiments focus on the preprocessing investigation by comparing preprocessed datasets with raw ones; (3) \textit{Module Comparison}: We evaluate the effectiveness of different feature selection and synthesis modules by reconstructing algorithms using them.

\subsection{Datasets} 
\label{subsec: data}
To make a comprehensive comparison, we would expect datasets to (1) vary in record and domain size, and attribute distribution to better show method performance on datasets of various complexity, and (2) differ in the proportion of numerical and categorical attributes to assess the preprocessing methods. Therefore, we choose five datasets used in previous work~\cite{liu2023generating, mckenna2022aim, zhang2021privsyn, kotelnikov2023tabddpm} as our datasets, which are ACSincome, ACSemploy, Bank, Higgs-small and Loan. \Rtwo{The brief information about these datasets is provided in \Cref{info:datasets}.} 

\Rone{
Furthermore, in \Cref{fig: datasets}, we plot the heat maps on absolute values of pairwise correlations across different datasets. Darker colors in the heat map mean higher correlation. We also plot the histograms of the attribute distribution's Shannon entropy, defined as 
$H(X) \;=\; -\sum_{i=1}^{k} p_i \,\ln p_i$.
The larger the entropy, the more informative and complex the distribution is. From \Cref{fig: datasets}, we can observe that ACSincome and ACSemploy have stronger correlations between attributes, while attribute distributions in the other three datasets have higher entropy, indicating greater complexity.
}

{
\setlength{\abovecaptionskip}{3pt}
\begin{figure}[t]
  \centering
  \includegraphics[width=0.75\columnwidth]{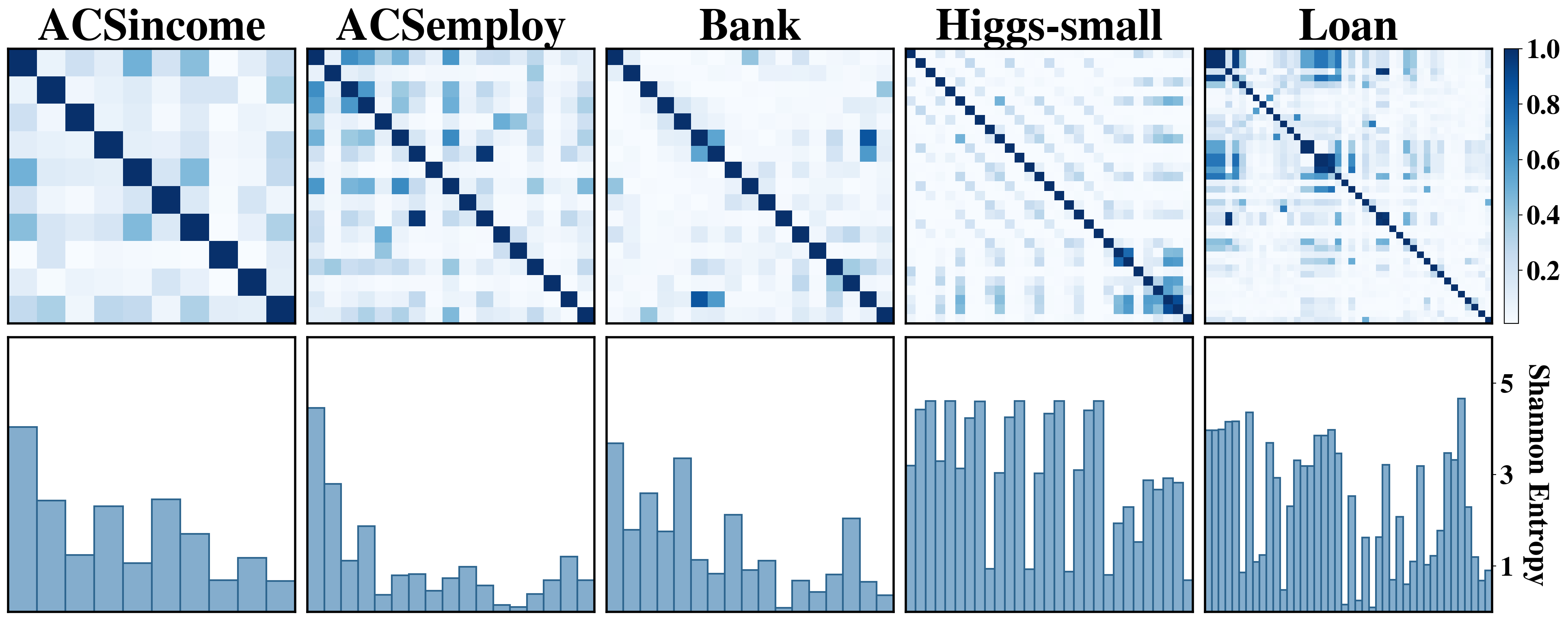}
    \caption{\Rone{Heat map of pairwise absolute correlations and histogram of attribute distribution's Shannon entropy.}}
    \label{fig: datasets}
\vspace{-4mm}
\end{figure}
}

\subsection{Implementations} 
\Rtwo{In our experiments, we consider 
\href{https://github.com/agl-c/deid2_dpsyn}{\uline{\privsyn}}, 
\href{https://github.com/caicre/PrivMRF}{\uline{\privmrf}}, 
\href{https://github.com/amazon-science/relaxed-adaptive-projection}{\uline{\rapp}}, 
\href{https://github.com/ryan112358/private-pgm}{\uline{\aim}},
\href{https://github.com/giusevtr/private_gsd}{\uline{\gsd}}, 
\href{https://github.com/terranceliu/iterative-dp?tab=readme-ov-file}{\uline{\gem}}, 
\href{https://github.com/ParkLabML/DP-MERF}{\uline{\merf}}, 
and \href{https://github.com/yandex-research/tab-ddpm}{\uline{\ddpm}}.
}
We do not include \rap in the experiments because \rapp directly improves upon it. Moreover, notice that \gsd is a synthesis algorithm; we use its one-shot 2-way marginal version in the overall evaluation, which is also used in their original work, and pay more attention to its performance in module comparison. Finally, we train \ddpm under DP-GSD by \textit{Opacus}~\cite{opacus}. We repeat all evaluations five times and report the average results. The DP parameter $\delta$ is set to be $10^{-5}$ by. We, by default, employ uniform binning and rare category merging preprocessing methods. The detailed hyperparameter settings of the studied methods, including preprocessing algorithms, are provided in the appendix of our full paper~\cite{chen2025benchmarking}. 

\Rtwo{
We keep our implementation of these methods aligned with their released versions. \privsyn and \aim are executed on CPU, while the other methods are executed with GPU. This is because \privsyn and \aim are designed for CPU and executing them on GPU may not be able to bring better performance. For instance, PrivSyn adjusts data in cliques one by one, which does not require high-dimensional computation. We also try \aim (PGM) on GPU (also with the officially released code from its authors) and compare the results with those from CPU. A brief result is shown in \Cref{tab: gpu res}. We can observe that there is no significant improvement in algorithm performance after executing on GPU.
}

\subsection{Evaluation Metrics}  
Various evaluation approaches have been proposed~\cite{zhang2021privsyn, mckenna2022aim, du2024towards}. However, most current works focus on comparing synthesis utility, or how similar synthetic data is to real data. While utility is important, algorithm efficiency is also a critical factor in algorithm selection. Therefore, we mainly consider the following metrics.

\vspace{0.5mm}
\noindent \textbf{Machine learning efficacy (higher is better)}. A widely accepted metric for evaluating generated data is training downstream machine learning (ML) models on the generated data and assessing their performance on test data. The previous works~\cite{zhang2021privsyn, kotelnikov2023tabddpm, du2024towards} typically select one or more ML models as downstream tasks.
It’s important to note that a large number of models do not necessarily lead to a fairer evaluation. Generally, simpler models have weaker data-fitting capabilities, which cannot reach fair conclusions. In our comparison, we therefore selected four deep learning models known for strong performance across various datasets: MLP, CatBoost, XGBoost, and Random Forest. We report the average F1 score on held-out test data as our metric value, and other related metrics, such as AUC and accuracy, are provided in the appendix of our full paper~\cite{chen2025benchmarking}. 

Another justification is that we use the test dataset instead of the training dataset for evaluation for this and the remaining metrics. While both approaches have been employed in previous works, we opt to use test data for evaluation due to the belief that it better reflects an algorithm's generalization ability.

{
\setlength{\abovecaptionskip}{3pt}
\setlength{\belowcaptionskip}{0pt}
\begin{table}[t]
\caption{\Rtwo{Results of AIM (PGM) on different hardware (CPU and GPU). These results are obtained on ACSincome dataset.}}
\label{tab: gpu res}
\resizebox{0.6\textwidth}{!}{
\begin{tabular}{lccccc}
\toprule
{$\varepsilon$} & {Method} & MLE$\uparrow$ & Query Err.$\downarrow$ & Fid Err.$\downarrow$ & Time \\
\midrule
\multirow{2}{*}{0.2} & AIM (CPU) & 0.77 & 0.0016 & 0.093 & 2.0 \text{min} \\
                     & AIM (GPU) & 0.75 & 0.0024 & 0.105 & 2.5 \text{min} \\
\midrule
\multirow{2}{*}{1.0} & AIM (CPU) & 0.78 & 0.0011 & 0.064 & 6.3 \text{min} \\
                     & AIM (GPU) & 0.78 & 0.0011 & 0.064 & 37.0 \text{min} \\
\bottomrule

\end{tabular}
}
\end{table}
}

\vspace{0.5mm}
\noindent \textbf{Query Error (lower is better)}. Making queries~\cite{chatfield2018introduction} is a commonly used data analysis technique, which can also be conducted to measure relatively high-dimensional similarity due to its high efficiency. Here, we consider using the 3-way marginal query method employed by Du et al. and McKenna et al.~\cite{du2024towards, mckenna2019graphical}, which utilizes the statistical $\ell_1$ error of frequency query results to reflect the magnitude of the error. Formally, the query error can be expressed as, $\mathbb{E}_{r \in R} \left| q_r(D_{syn}) - q_r(D_{test}) \right|,$
where $q_r$ refers to the query function, which is a combination of range query (for numerical attributes) and point query (for categorical attributes). $\mathbb{E}$ is the mathematical expectation, and $R$ refers to the set of all 3-way marginals. \Rthree{Moreover, due to complex real applications on synthetic data, we also worry that simple frequency queries, even though fundamental, may be insufficient to cover them. Therefore, we also consider \textbf{Conditional Query Error} to simulate \texttt{Join} and \texttt{GroupBy} in SQL (fix one attribute as the key column and query the frequency of the other attribute). This metric serves as a supplementary evaluation reference, providing comprehensive comparisons.}

\vspace{0.5mm}
\noindent \textbf{Fidelity Error (lower is better)}. Marginal Fidelity is precise in evaluating low-dimensional similarity, such as average 2-way marginal discrepancy. Total variation distance (TVD) can be used for such measurement, which has also been utilized in some studies~\cite{tao2022benchmarkingdifferentiallyprivatesynthetic, tran2024differentially}. We define the TVD as
$\frac{1}{2}\sum_{1 \leq i \leq j \leq d} \left| M^{\text{syn}}_{i,j} - M^{\text{test}}_{i,j}\right|,$
where $M^{\text{syn}}_{i,j}$ and $M^{\text{test}}_{i,j}$ are the real 2-way marginals determined by the synthetic dataset and test dataset, respectively. 

\vspace{0.5mm}
\noindent \textbf{Running Time (lower is better)}. A straightforward measurement of algorithm efficiency is the execution time when generating the same amount of data. The running time does not include the preprocessing step and only counts the time spent in feature selection and data synthesis modules.  

\vspace{0.4mm}
\Rtwo{In addition to our primary evaluation criteria, there are many other widely applied metrics for tabular data, such as Wasserstein distance~\cite{peyre2019computational} and Cramer’s V measure~\cite{cramer1999mathematical}. We do not include these metrics in our evaluation mainly because of their feasibility in complex cases and overlap with existing metrics. For example, Wasserstein distance can be infeasible in terms of computation time when dealing with high-dimensional and complex data. Moreover, Cramer’s V measure and other correlation measurements cannot provide deterministic information about data distributions. In other words, even if two distributions are different, they can have similar correlation results. Furthermore, correlation information can be covered by fidelity error, which is based on a precise measurement of total distribution. We will further validate this in the appendix of our full paper~\cite{chen2025benchmarking}.}

{
\setlength{\abovecaptionskip}{3pt}
\setlength{\belowcaptionskip}{0pt}
\begin{table*}[!t]
\footnotesize
    \centering
    \setlength{\tabcolsep}{4.5pt}
    \caption{Overall utility of synthetic data under different methods. The results with the best performance are highlighted in bold (due to the limited number of digits displayed, some data may appear equal in the table, but there is actually an order in their values). Ground Truth is obtained by comparing real data with test data.}
    \resizebox{0.99\textwidth}{!}{
    \begin{tabular}{l|ccc|ccc|ccc|ccc|ccc}
    \toprule 
        \textbf{Dataset} & \multicolumn{3}{c|}{\textbf{ACSincome}} & \multicolumn{3}{c|}{\textbf{ACSemploy}} & \multicolumn{3}{c|}{\textbf{Bank}} & \multicolumn{3}{c|}{\textbf{Higgs-small}} & \multicolumn{3}{c}{\textbf{Loan}}\\ 
        \cmidrule{1-16}
        \textbf{ML Efficacy $\uparrow$} & $\varepsilon = 0.2$ & $\varepsilon = 1$ & $\varepsilon = 5$ & $\varepsilon = 0.2$ & $\varepsilon = 1$ & $\varepsilon = 5$ & $\varepsilon = 0.2$ & $\varepsilon = 1$ & $\varepsilon = 5$ & $\varepsilon = 0.2$ & $\varepsilon = 1$ & $\varepsilon = 5$ & $\varepsilon = 0.2$ & $\varepsilon = 1$ & $\varepsilon = 5$ \\ 
        \midrule
        \privsyn     & $0.43$ & $0.39$ & $0.42$ & $0.40$ & $0.43$ & $0.39$ & $0.47$ & $0.47$ & $0.47$ & $0.40$ & $0.43$ & $0.43$ & $0.25$ & $0.26$ & $0.26$ \\
        \privmrf     & $0.73$ & $0.78$ & $0.78$ & $0.72$ & $0.80$ & $0.81$ & $0.62$ & $0.69$ & $\textbf{0.71}$ & $0.50$ & $0.64$ & $0.64$ & $\textbf{0.52}$ & $\textbf{0.52}$ & $\textbf{0.52}$ \\
        \rapp        & $0.66$ & $0.73$ & $0.77$ & $0.74$ & $0.77$ & $0.80$ & $0.65$ & $0.69$ & $0.67$ & $0.52$ & $0.53$ & $0.54$ & $0.45$ & $0.42$ & $0.43$ \\
        \aim         & $\textbf{0.76}$ & $\textbf{0.78}$ & $\textbf{0.78}$ & $\textbf{0.78}$ & $\textbf{0.80}$ & $\textbf{0.81}$ & $\textbf{0.67}$ & $\textbf{0.71}$ & $0.71$ & $\textbf{0.63}$ & $\textbf{0.64}$ & $\textbf{0.67}$ & $0.52$ & $0.52$ & $0.52$ \\
        \gsd         & $0.76$ & $0.77$ & $0.77$ & $0.72$ & $0.73$ & $0.72$ & $0.47$ & $0.48$ & $0.49$ & $0.48$ & $0.47$ & $0.49$ & $0.25$ & $0.24$ & $0.25$ \\
        \gem         & $0.70$ & $0.68$ & $0.66$ & $0.67$ & $0.70$ & $0.69$ & $0.51$ & $0.56$ & $0.53$ & $0.51$ & $0.52$ & $0.52$ & $0.50$ & $0.49$ & $0.51$ \\
        \merf        & $0.65$ & $0.67$ & $0.71$ & $0.58$ & $0.71$ & $0.66$ & $0.60$ & $0.57$ & $0.55$ & $0.53$ & $0.56$ & $0.57$ & $0.32$ & $0.18$ & $0.18$ \\
        \ddpm        & $0.41$ & $0.41$ & $0.39$ & $0.48$ & $0.42$ & $0.51$ & $0.47$ & $0.47$ & $0.47$ & $0.36$ & $0.35$ & $0.34$ & $0.24$ & $0.24$ & $0.24$ \\
        \hline
        \rowcolor{gray!15} \textbf{Ground Truth} & $0.79$ & $0.79$ & $0.79$ & $0.81$ & $0.81$ & $0.81$ & $0.76$ & $0.76$ & $0.76$ & $0.72$ & $0.72$ & $0.72$ & $0.54$ & $0.54$ & $0.54$ \\
    \bottomrule
    \toprule 
        \textbf{Query Error $\downarrow$} & $\varepsilon = 0.2$ & $\varepsilon = 1$ & $\varepsilon = 5$ & $\varepsilon = 0.2$ & $\varepsilon = 1$ & $\varepsilon = 5$ & $\varepsilon = 0.2$ & $\varepsilon = 1$ & $\varepsilon = 5$ & $\varepsilon = 0.2$ & $\varepsilon = 1$ & $\varepsilon = 5$ & $\varepsilon = 0.2$ & $\varepsilon = 1$ & $\varepsilon = 5$ \\ 
        \midrule
        \privsyn     & $0.003$ & $0.002$ & $0.002$ & $0.008$ & $0.004$ & $0.004$ & $0.007$ & $0.004$ & $0.003$ & $0.009$ & $0.004$ & $0.003$ & $0.006$ & $0.005$ & $0.004$\\
        \privmrf     & $0.002$ & $0.001$ & $0.001$ & $0.004$ & $0.002$ & $0.002$ & $\textbf{0.005}$ & $0.003$ & $0.003$ & $0.005$ & $0.005$ & $0.003$ & $\textbf{0.005}$ & $\textbf{0.005}$ & $\textbf{0.004}$\\
        \rapp        & $0.019$ & $0.005$ & $0.003$ & $0.029$ & $0.009$ & $0.003$ & $0.014$ & $0.006$ & $0.005$ & $0.035$ & $0.029$ & $0.028$ & $0.020$ & $0.014$ & $0.011$ \\
        \aim         & $\textbf{0.002}$ & $\textbf{0.001}$ & $\textbf{0.001}$ & $\textbf{0.004}$ & $\textbf{0.002}$ & $\textbf{0.001}$ & $0.007$ & $\textbf{0.002}$ & $\textbf{0.002}$ & $\textbf{0.005}$ & $\textbf{0.003}$ & $\textbf{0.003}$ & $0.005$ & $0.005$ & $0.004$ \\
        \gsd         & $0.004$ & $0.003$ & $0.002$ & $0.026$ & $0.026$ & $0.026$ & $0.044$ & $0.044$ & $0.043$ & $0.044$ & $0.044$ & $0.044$ & $0.038$ & $0.037$ & $0.036$\\
        \gem         & $0.014$ & $0.017$ & $0.016$ & $0.010$ & $0.006$ & $0.006$ & $0.118$ & $0.021$ & $0.022$ & $0.066$ & $0.065$ & $0.065$ & $0.030$ & $0.030$ & $0.029$ \\
        \merf        & $0.019$ & $0.018$ & $0.024$ & $0.039$ & $0.037$ & $0.036$ & $0.038$ & $0.035$ & $0.036$ & $0.039$ & $0.035$ & $0.034$ & $0.006$ & $0.006$ & $0.006$\\
        \ddpm        & $0.066$ & $0.064$ & $0.060$ & $0.088$ & $0.067$ & $0.079$ & $0.074$ & $0.071$ & $0.088$ & $0.106$ & $0.106$ & $0.107$ & $0.067$ & $0.066$ & $0.070$ \\
        \hline
        \rowcolor{gray!15} \textbf{Ground Truth} & $0.001$ & $0.001$ & $0.001$ & $0.001$ & $0.001$ & $0.001$ & $0.001$ & $0.001$ & $0.001$ & $0.001$ & $0.001$ & $0.001$ & $0.001$ & $0.001$ & $0.001$ \\
    \bottomrule
    \toprule
        \textbf{Fidelity Error $\downarrow$} & $\varepsilon = 0.2$ & $\varepsilon = 1$ & $\varepsilon = 5$ & $\varepsilon = 0.2$ & $\varepsilon = 1$ & $\varepsilon = 5$ & $\varepsilon = 0.2$ & $\varepsilon = 1$ & $\varepsilon = 5$ & $\varepsilon = 0.2$ & $\varepsilon = 1$ & $\varepsilon = 5$ & $\varepsilon = 0.2$ & $\varepsilon = 1$ & $\varepsilon = 5$ \\ 
        \midrule
        \privsyn     & $0.15$ & $0.12$ & $0.12$ & $0.12$ & $0.09$ & $0.09$ & $0.24$ & $0.13$ & $0.12$ & $0.29$ & $0.20$ & $0.15$ & $0.34$ & $0.35$ & $0.36$\\
        \privmrf     & $0.11$ & $0.07$ & $0.05$ & $0.07$ & $0.04$ & $0.03$ & $0.13$ & $\textbf{0.07}$ & $\textbf{0.06}$ & $0.21$ & $\textbf{0.16}$ & $0.16$ & $\textbf{0.31}$ & $\textbf{0.24}$ & $\textbf{0.23}$ \\
        \rapp        & $0.52$ & $0.24$ & $0.19$ & $0.30$ & $0.13$ & $0.07$ & $0.43$ & $0.36$ & $0.36$ & $0.69$ & $0.65$ & $0.65$ & $0.66$ & $0.58$ & $0.55$ \\
        \aim         & $\textbf{0.09}$ & $\textbf{0.06}$ & $\textbf{0.05}$ & $\textbf{0.05}$ & $\textbf{0.03}$ & $\textbf{0.02}$ & $\textbf{0.11}$ & $0.09$ & $0.09$ & $\textbf{0.19}$ & $0.17$ & $\textbf{0.14}$ & $0.35$ & $0.32$ & $0.29$ \\
        \gsd         & $0.23$ & $0.21$ & $0.20$ & $0.22$ & $0.22$ & $0.22$ & $0.52$ & $0.52$ & $0.52$ & $0.65$ & $0.65$ & $0.65$ & $0.67$ & $0.66$ & $0.66$\\
        \gem         & $0.26$ & $0.28$ & $0.27$ & $0.15$ & $0.08$ & $0.09$ & $0.76$ & $0.21$ & $0.23$ & $0.57$ & $0.57$ & $0.36$ & $0.53$ & $0.52$ & $0.52$\\
        \merf        & $0.52$ & $0.51$ & $0.50$ & $0.34$ & $0.34$ & $0.32$ & $0.47$ & $0.48$ & $0.48$ & $0.60$ & $0.56$ & $0.54$ & $0.91$ & $0.92$ & $0.93$\\
        \ddpm        & $0.78$ & $0.77$ & $0.71$ & $0.60$ & $0.51$ & $0.57$ & $0.79$ & $0.78$ & $0.82$ & $0.70$ & $0.81$ & $0.88$ & $0.95$ & $0.95$ & $0.94$\\
        \hline
        \rowcolor{gray!15} \textbf{Ground Truth} & $0.05$ & $0.05$ & $0.05$ & $0.02$ & $0.02$ & $0.02$ & $0.03$ & $0.03$ & $0.03$ & $0.12$ & $0.12$ & $0.12$ & $0.10$ & $0.10$ & $0.10$ \\
    \bottomrule
    \end{tabular}
    }
    \label{general result}
\end{table*}
}

{
\setlength{\abovecaptionskip}{8pt}
\begin{figure*}[t]
    \centering
    \begin{minipage}{\textwidth}
    \hspace*{0.01\textwidth}
    \begin{subfigure}[b]{0.23\textwidth}
        \centering
        \includegraphics[width=\textwidth]{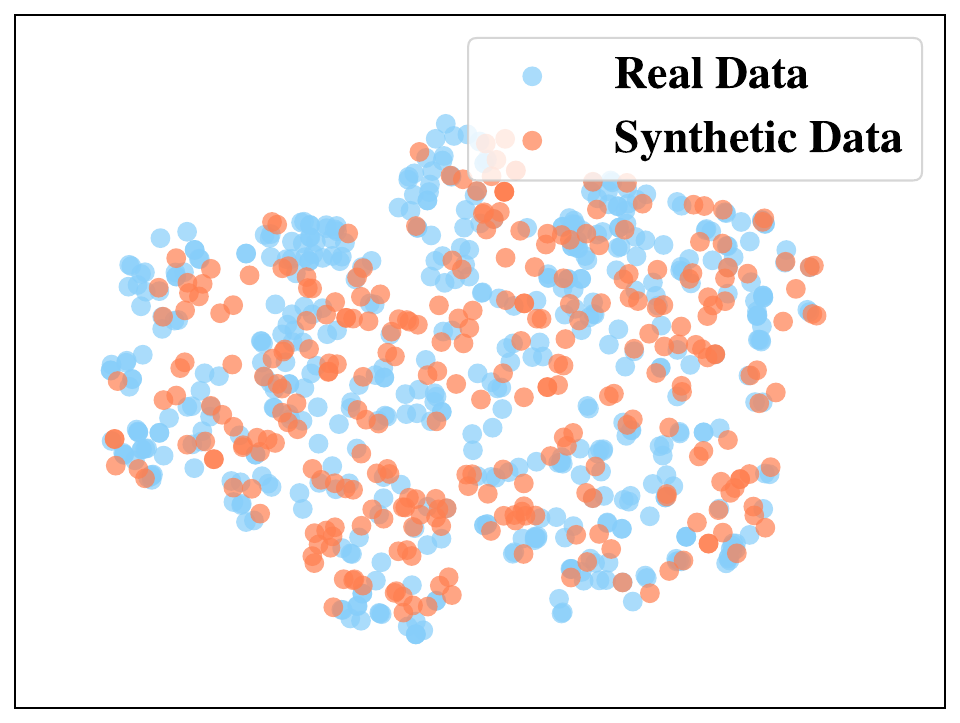}
        \caption{PrivSyn}
        \label{fig0:privsyn}
    \end{subfigure}
    \hspace{0.012\textwidth}
    \begin{subfigure}[b]{0.23\textwidth}
        \centering
        \includegraphics[width=\textwidth]{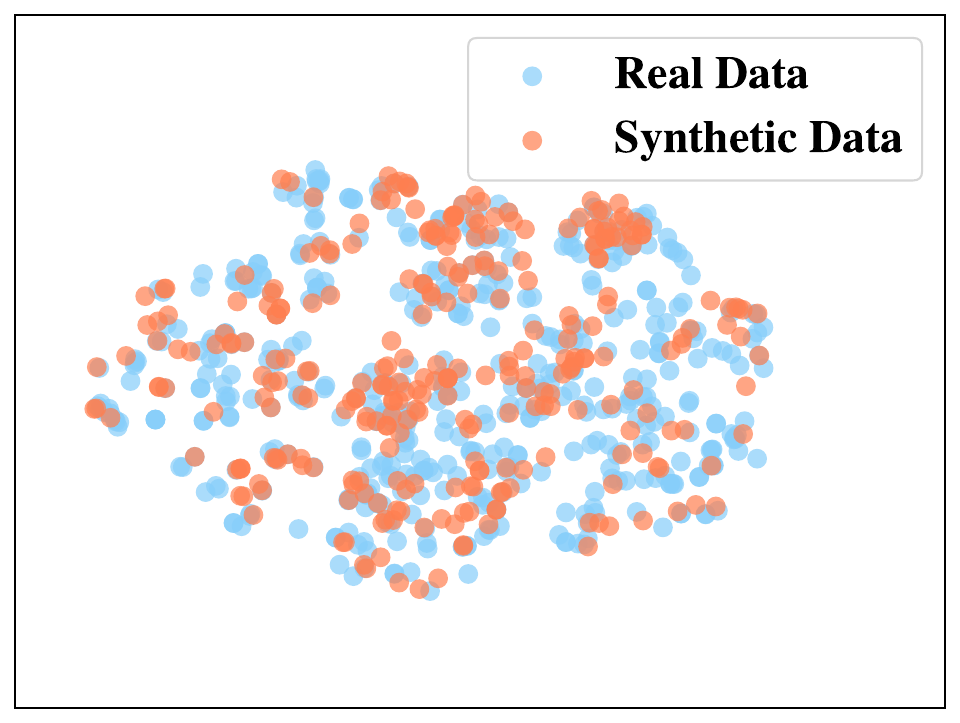}
        \caption{PrivMRF}
        \label{fig0:mrf}
    \end{subfigure}
    \hspace{0.012\textwidth}
    \begin{subfigure}[b]{0.23\textwidth}
        \centering
        \includegraphics[width=\textwidth]{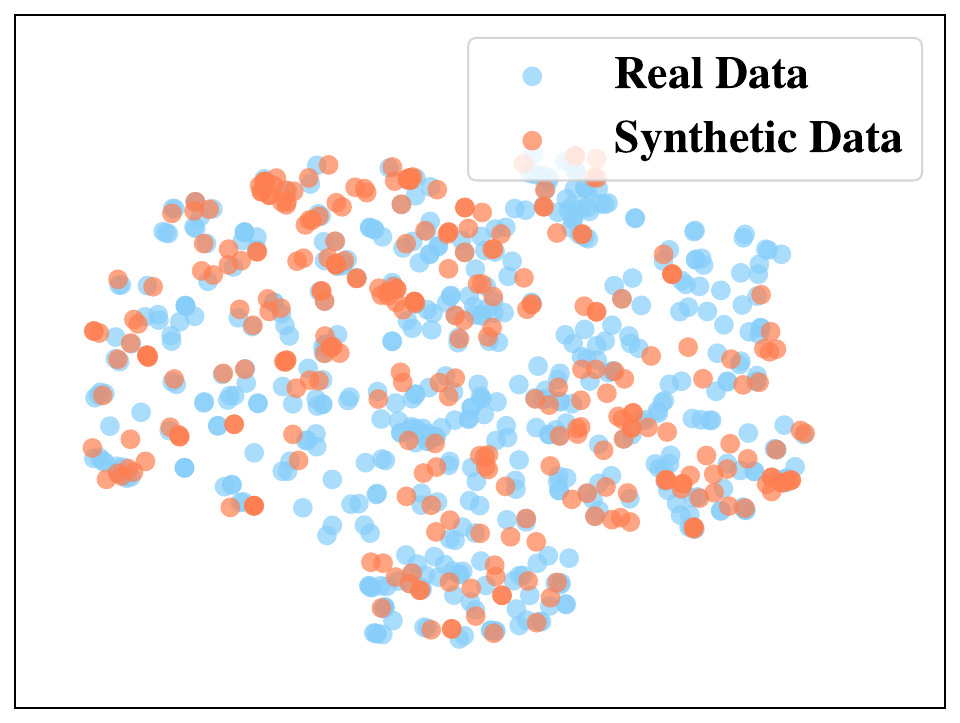}
        \caption{RAP++}
        \label{fig0:rap}
    \end{subfigure}
    \hspace{0.012\textwidth}
    \begin{subfigure}[b]{0.23\textwidth}
        \centering
        \includegraphics[width=\textwidth]{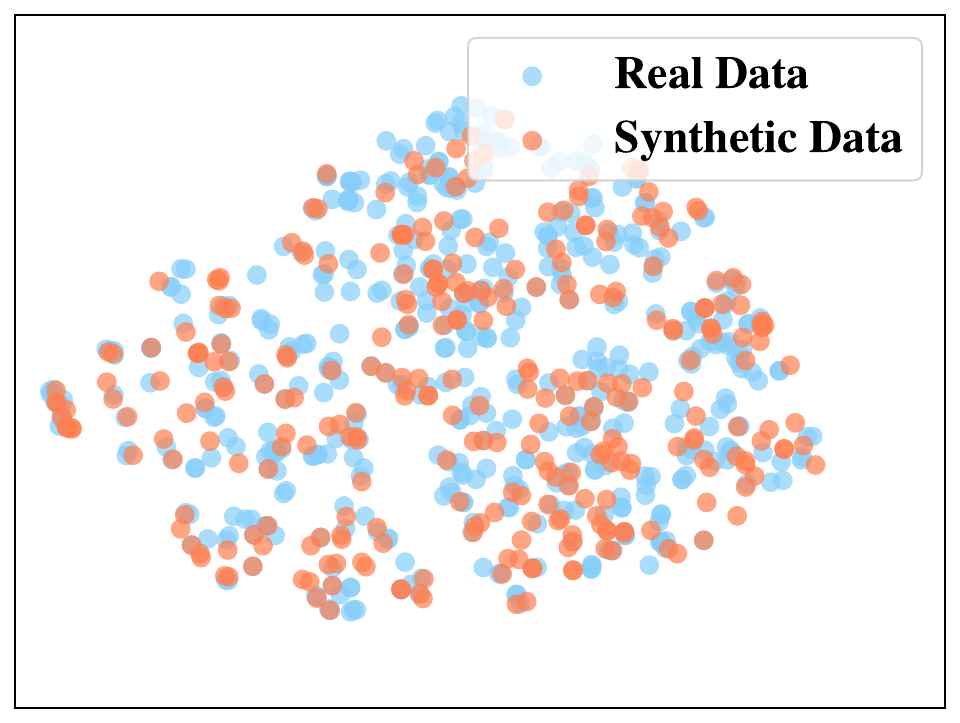}
        \caption{AIM}
        \label{fig0:aim}
    \end{subfigure}
    \\
    \hspace*{0.01\textwidth}
    \begin{subfigure}[b]{0.23\textwidth}
        \centering
        \includegraphics[width=\textwidth]{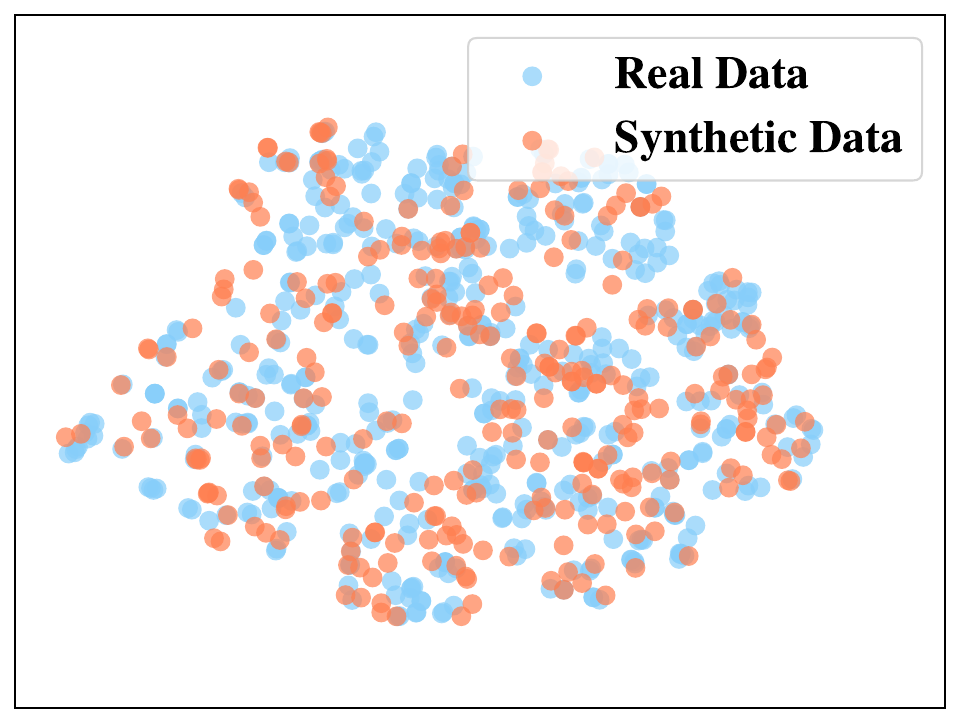}
        \caption{Privtae-GSD}
        \label{fig0:gsd}
    \end{subfigure}
    \hspace{0.012\textwidth}
    \begin{subfigure}[b]{0.23\textwidth}
        \centering
        \includegraphics[width=\textwidth]{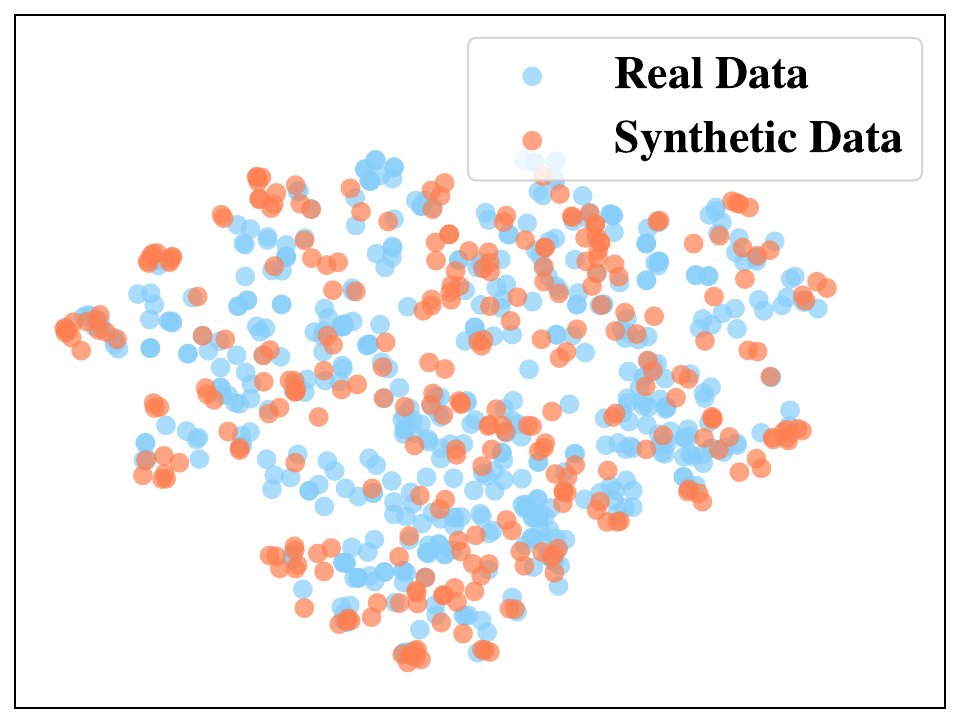}
        \caption{GEM}
        \label{fig0:gem}
    \end{subfigure}
    \hspace{0.012\textwidth}
    \begin{subfigure}[b]{0.23\textwidth}
        \centering
        \includegraphics[width=\textwidth]{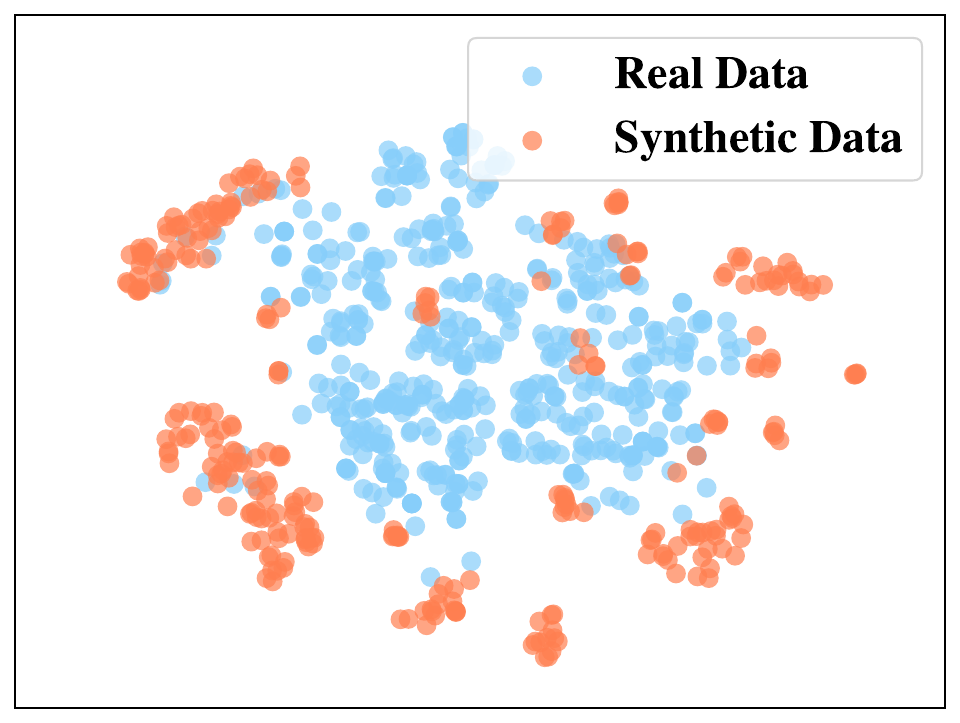}
        \caption{DP-MERF}
        \label{fig0:merf}
    \end{subfigure}
    \hspace{0.012\textwidth}
    \begin{subfigure}[b]{0.23\textwidth}
        \centering
        \includegraphics[width=\textwidth]{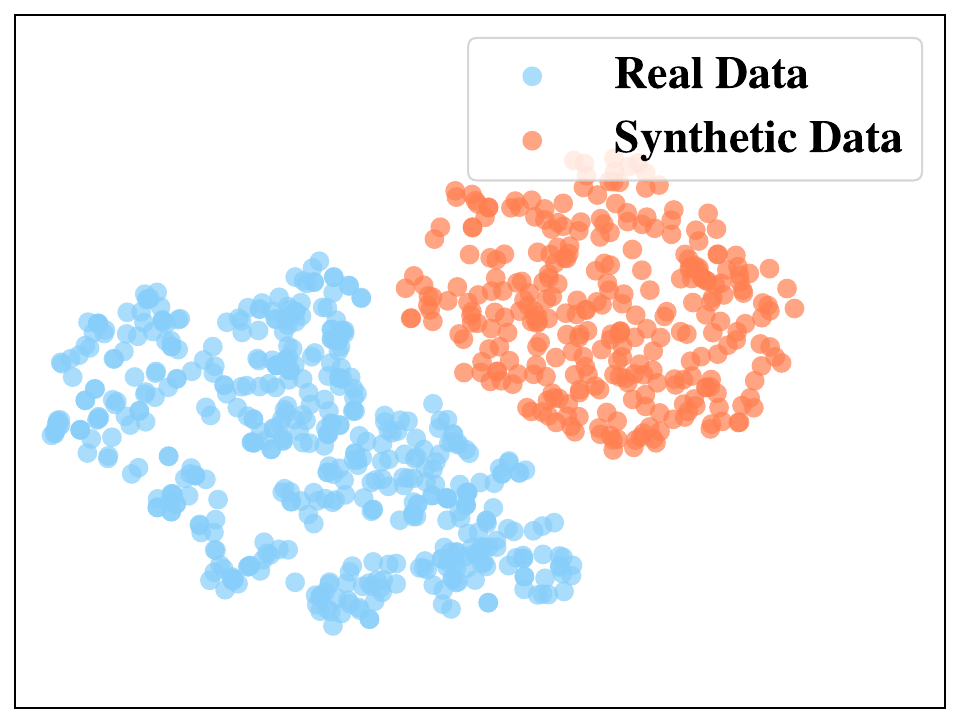}
        \caption{TabDDPM}
        \label{fig0:ddpm}
    \end{subfigure}    
    \end{minipage}
    \vspace{-2mm}
    \caption{T-SNE scatter plots of synthesis results on Bank dataset under $\varepsilon = 1.0$}
    \label{fig0: exp}
    \vspace{-2mm}
\end{figure*}
}

\section{Experimental Results}
\label{sec: exp}

\subsection{Overall Evaluation} 
\label{overall exp}
\noindent \textbf{Utility Comparison.} We provide detailed results on machine learning efficacy, query error, and fidelity error in \cref{general result}, and show some brief results of conditional query error in \cref{cond query}. To make the result more straightforward, we utilize the t-SNE~\cite{van2008visualizing}, a dimensionality reduction technique, to visualize the synthesis results of all methods on the Bank dataset in \cref{fig0: exp}. By reducing the dimensionality of the dataset with t-SNE technique and plotting the scatter distribution, these visualizations reveal the distribution overlap between synthetic and
real data, demonstrating synthesizers' ability to generate data resembling the original. 

The first clear conclusion is that \privmrf and \aim achieve the best utility metrics among all methods. Another straightforward finding is that statistical methods generally perform better than deep learning approaches. Both \aim\ and \privmrf rely on graphical models, and the quantitative evaluations of \gem, \merf, and \ddpm are worse than most other statistical methods in some metrics, such as \rapp and \privsyn in fidelity error.

\vspace{0.5mm}
\noindent \textbf{Utility \& Efficiency Trade-off.} We present the relationship between utility and efficiency on different algorithms in \cref{fig: trade}. Although \aim and \privmrf exhibit excellent utility performance, they require more execution time, representing a trade-off in utility. \privmrf is slightly more efficient than \aim because it involves an initialization step, reducing the number of selection rounds. Additionally, all deep learning-based methods, despite having lower generation utility, demonstrate high efficiency. This is partly because some statistical methods run on CPU and also because they involve large-scale computations.

\vspace{0.5mm}
\noindent \textbf{\ddpm has a weak performance in synthesis utility.} Among all the methods, \ddpm performs poorly in all three dimensions, and in \cref{fig0: exp}, the visualization of the synthesis result does not match the real data distribution. We track the training loss with DP-SGD and compare it with the loss without DP-SGD. In \cref{fig1:sub1}, we can observe that the loss under DP-SGD does not converge well when training. This raises our suspicion of DP-SGD's efficiency. However, the diffusion model still demonstrates excellent performance when generating images~\cite{li2024privimage, dockhorn2023differentiallyprivatediffusionmodels} under DP-SGD. Therefore, we hypothesize that the failure of \ddpm is due to its incompatibility with tabular data under a DP-SGD framework. Unlike image data, which is characterized by dense and repetitive features (e.g., visual similarity among images of the same type), tabular data exhibits sparse and highly divergent characteristics (e.g., unique relationships between attributes), which present significant convergence challenges for DP-SGD.

{
\setlength{\abovecaptionskip}{5pt}
\setlength{\belowcaptionskip}{3pt}
\begin{figure*}[t]
    \centering
    \begin{subfigure}[b]{0.32\textwidth}
        \centering
        \includegraphics[width=\textwidth]{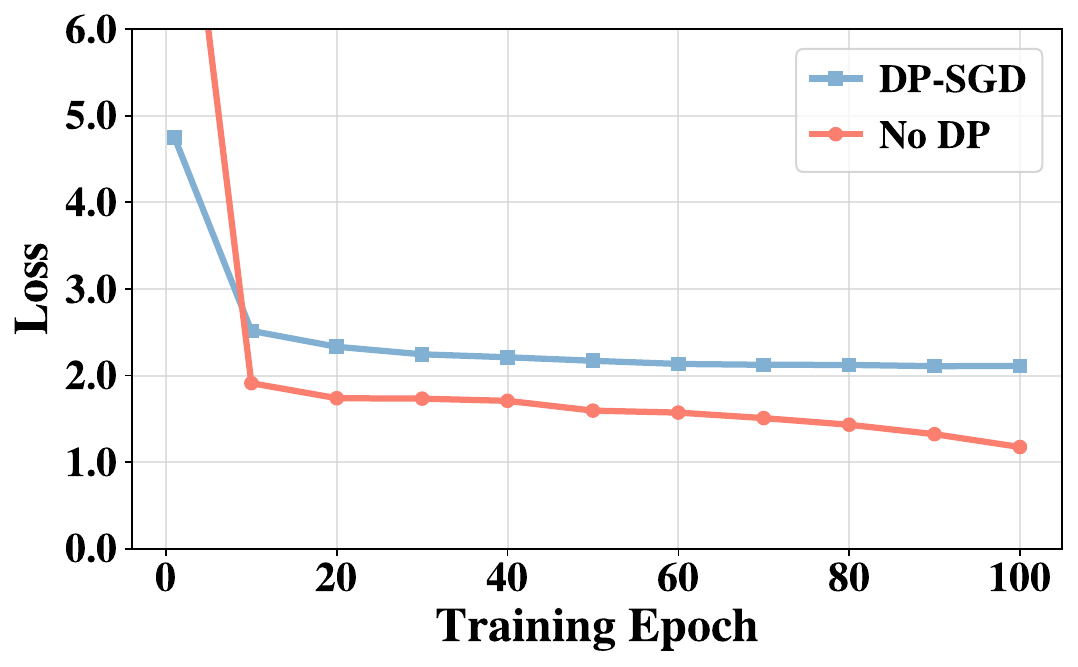}
        \caption{Training loss of \ddpm with and without DP on Bank dataset under $\varepsilon=1.0$.}
        \label{fig1:sub1}
    \end{subfigure}
    \hspace{0.005\textwidth}
    \begin{subfigure}[b]{0.32\textwidth}
        \centering
        \includegraphics[width=\textwidth]{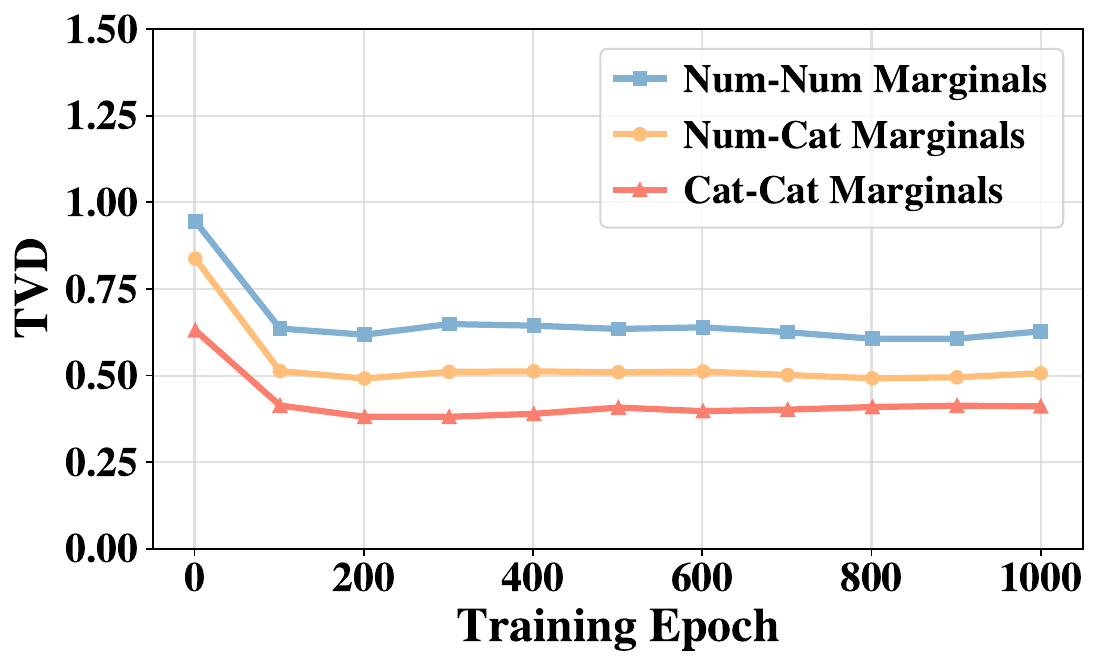}
        \caption{TVD of marginals between generated data and real data when training \merf.}
        \label{fig1:sub2}
    \end{subfigure}
    \hspace{0.005\textwidth}
    \begin{subfigure}[b]{0.32\textwidth}
        \centering
        \includegraphics[width=\textwidth]{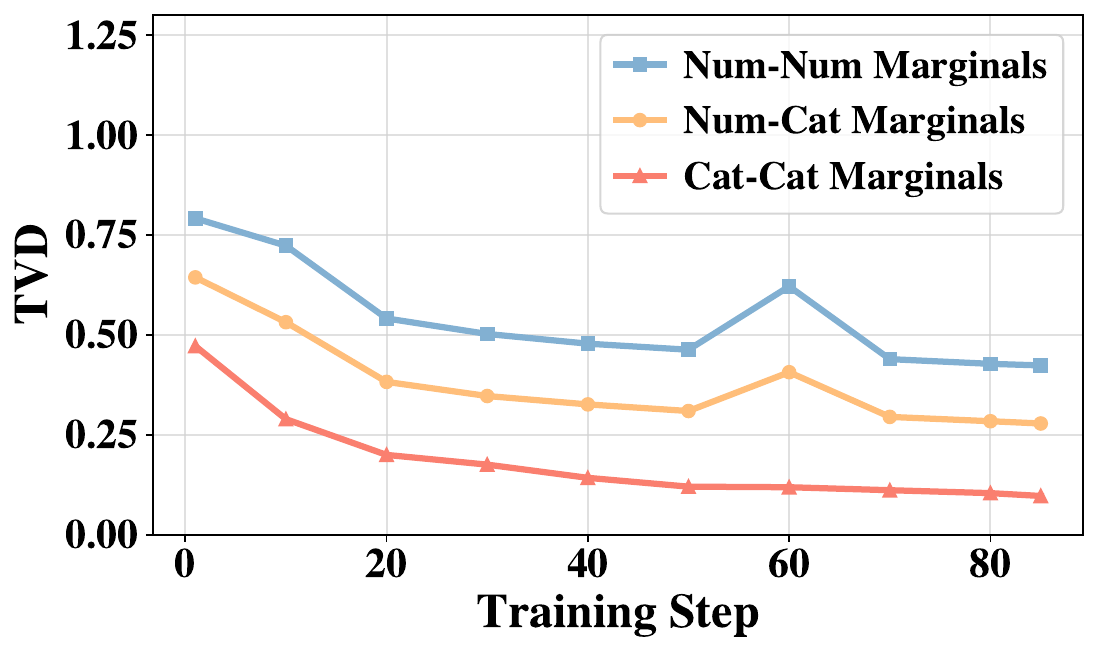}
        \caption{TVD of marginals between generated data and real data when training \gem.}
        \label{fig1:sub3}
    \end{subfigure}
    \vspace{-2mm}
    \caption{Figures for analyzing different methods in \cref{overall exp}}
    \label{fig1: exp}
    \vspace{-3mm}
\end{figure*}
}

{
\setlength{\abovecaptionskip}{3pt}
\setlength{\belowcaptionskip}{0pt}
\begin{table}[t]
\footnotesize
    \centering
    \caption{\Rthree{Conditional query errors of different methods under $\varepsilon=1.0$.}}
    \label{cond query}
    \resizebox{0.6\columnwidth}{!}{
    \begin{tabular}{l|ccccc}
    \toprule
    \textbf{Method} & \textbf{ACSincome} & \textbf{ACSemploy} & \textbf{Bank} & \textbf{Higgs-small} & \textbf{Loan} \\
    \midrule
    Privsyn & 0.019 & 0.035 & 0.033 & 0.029 & 0.058 \\
    PrivMRF & \textbf{0.012} & 0.013 & 0.026 & 0.026 & \textbf{0.053} \\
    RAP++ & 0.021 & 0.026 & 0.064 & 0.098 & 0.098 \\
    AIM & 0.013 & \textbf{0.011} & \textbf{0.022} & \textbf{0.025} & 0.057 \\
    Private-GSD & 0.016 & 0.014 & 0.039 & 0.110 & 0.093 \\
    GEM & 0.026 & 0.028 & 0.053 & 0.069 & 0.077 \\
    DP-MERF & 0.308 & 0.314 & 0.215 & 0.099 & 0.455 \\
    TabDDPM & 0.092 & 0.072 & 0.136 & 0.109 & 0.143 \\
    \bottomrule
    \end{tabular}
    }
    \vspace{-4mm}
\end{table}
}

{
\setlength{\abovecaptionskip}{3pt}
\setlength{\belowcaptionskip}{-5pt}
\begin{figure*}
    \centering
    \includegraphics[width=1.0\textwidth]{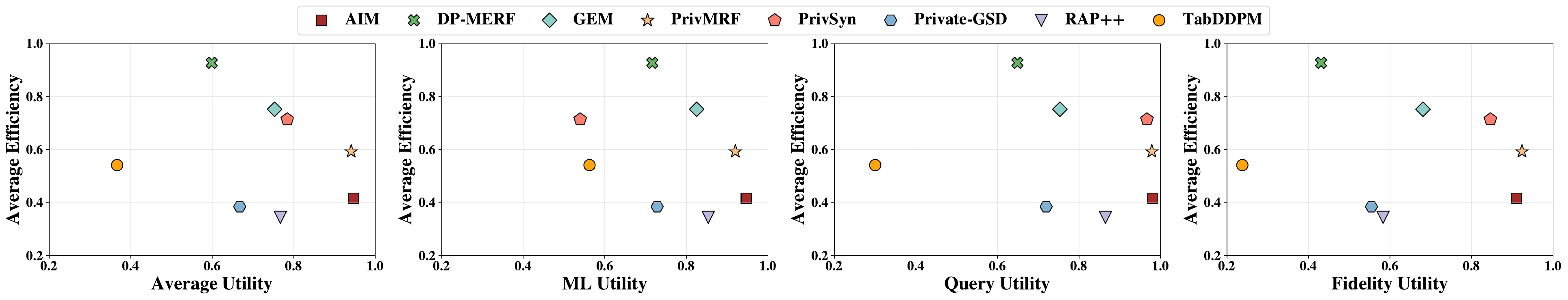}
    \vspace{-3.5mm}
    \caption{\Rone{Scaled utility and efficiency of different algorithms. Average utility is obtained by taking the average of ML efficacy, query error, and fidelity error after normalizing them to $[0,1]$, guaranteeing their equal contribution to the aggregated metric. Other metrics are directly obtained from the average value of normalized original metrics.}}
    \label{fig: trade}
\end{figure*}
}

\vspace{0.5mm}
\noindent \textbf{Two GAN-based methods, \merf and \gem, perform differently.} 
Even though \gem and \merf both use deep generative networks for synthesizing, compared to \gem, \merf performs poorly regarding query and fidelity error. One key reason lies in the construction of features. Firstly, random Fourier features embedding, used by \merf, is still approximating the joint distribution of numerical attributes, which inherently introduces approximation error. Furthermore, when dealing with categorical variables, \merf focuses solely on the marginal distributions between categorical variables and the label variable, which overlooks other marginals. While \gem employs an adaptive marginal selection strategy, continuously adjusting the fitting target.

To verify our hypothesis, we track the fitting errors of different categories of marginals in \cref{fig1:sub2} and \cref{fig1:sub3}. 
During the \merf training process, the total variation distance (TVD) of most marginals, except those for ``categorical-categorical'' marginals, does not effectively converge. This, to some extent, supports our thinking. By comparison, \gem achieves better convergence across marginals.
Moreover, in \cref{fig0:merf}, we observe that the data points generated by \merf concentrate in several areas, which we believe is caused by unbalanced feature construction. 
In contrast, as shown in \cref{fig0:gem}, the data distribution of \gem is better aligned with the target dataset.

\vspace{0.5mm} 
\noindent \textbf{\privsyn works poorly on machine learning efficacy.} 
Another finding is that \privsyn performs well on range query error and fidelity error metrics, but underperforms in terms of machine learning efficacy. This scenario aligns with our previous analysis in \cref{sec: syn}, which expresses the concern that \mtdtt{GUM} focuses more on local marginal cliques and may ignore global relevance. The experiments in \cref{divide exp} can also support this statement.

{
\setlength{\abovecaptionskip}{3pt}
\setlength{\belowcaptionskip}{-1pt}
\begin{figure*}[t]
    \centering
    \includegraphics[width=\textwidth]{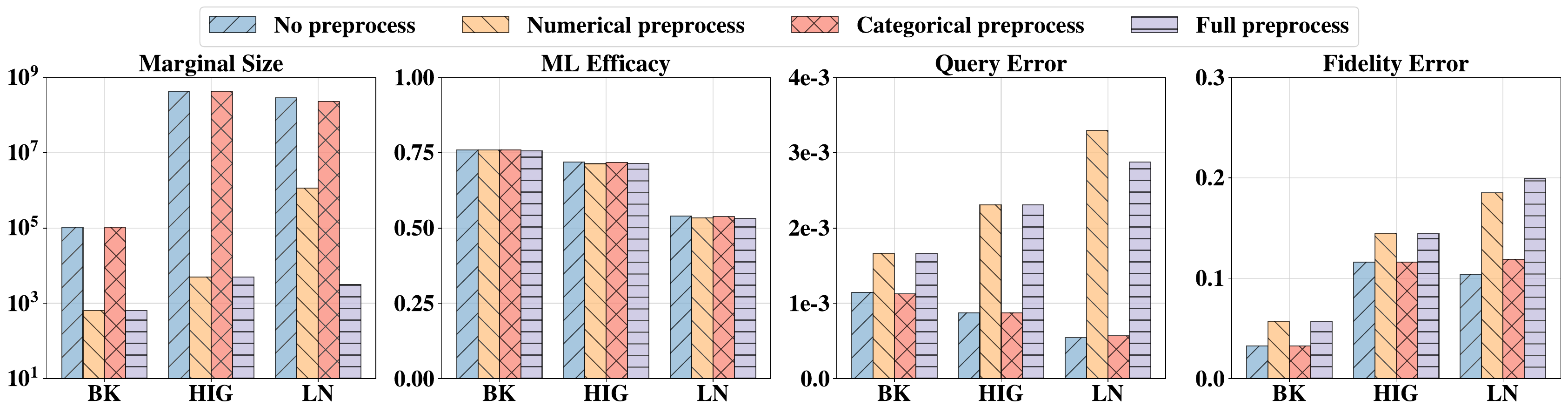}
    \caption{Metrics under different preprocessing settings. These metrics are obtained by comparing preprocessed raw data with test data. Numerical preprocessing means only conducting numerical discretization, and categorical preprocessing means only conducting categorical rare merging. By default, the results are under the setting that total privacy budget $\varepsilon = 1.0$.}
    \label{fig2: exp}
    \vspace{-1mm}
\end{figure*}
}

\subsection{Preprocessing Investigation}
\label{preprocess exp}
In this Section, we will explore the influence of preprocessing on algorithms' performance, and how different preprocessing strategies will influence the quality of synthesis data.

\vspace{0.5mm}
\noindent \textbf{Influence of Preprocessing on Dataset Information.} Directly comparing the algorithms' performances with and without preprocessing is difficult because running algorithms on raw datasets is often too time-consuming and computationally complex (e.g., some methods even require more than 24 hours to run on raw datasets). Therefore, we consider comparing the marginal sizes and utility metrics calculated on preprocessed and raw datasets. The detailed results are shown in \cref{fig2: exp}. A straightforward finding is that preprocessing decreases the complexity of data, with the introduction of only a small error. In \cref{fig2: exp}, the average marginal size significantly decreases after preprocessing (from $10^8$ to $10^3$ in Higgs-small and Loan datasets). Meanwhile, its negative influences on utility are small enough (change on query error $<$ 0.003, TVD $<$ 0.1). We infer this is because binning can preserve most numerical characteristics, and the low-frequency categorical values only contribute a small proportion of the overall correlation between attributes in the dataset.

{
\setlength{\abovecaptionskip}{3pt}
\setlength{\belowcaptionskip}{-2pt}
\begin{table}[t]
\footnotesize
    \centering
    \caption{\Rtwo{Different algorithms' performance on ACSincome with and without preprocessing under $\varepsilon = 1.0$. Here, ``Prep" means preprocessed data, and ``Raw" means raw data.}}
    \label{enforce pre}
    \resizebox{0.6\columnwidth}{!}{
    \begin{tabular}{l|cc|cc|cc|cc}
    \toprule
     \multirow{2}{*}{\textbf{Method}} & \multicolumn{2}{c|}{\textbf{ML Eff. $\uparrow$}} & \multicolumn{2}{c|}{\textbf{Query Err. $\downarrow$}} & \multicolumn{2}{c|}{\textbf{Fid Err. $\downarrow$}} & \multicolumn{2}{c}{\textbf{Time (min) $\downarrow$}}\\ 
     \cmidrule{2-9}
     & Prep & Raw & Prep & Raw & Prep & Raw & Preps & Raw \\
    \midrule
    \privsyn & 0.39 & 0.39 & 0.002 & 0.002 & 0.12 & 0.12 & 0.2 & 0.2 \\
    \privmrf & 0.78 & 0.78 & 0.001 & 0.001 & 0.06 & 0.07 & 1.5 & 0.9 \\
    \aim & 0.78 & 0.78 & 0.001 & 0.001 & 0.06 & 0.06 & 5.8 & 6.3 \\
    \rapp & 0.74 & 0.74 & 0.004 & 0.005 & 0.22 & 0.24 & 13.3 & 26.8 \\
    \gsd & 0.77 & 0.77 & 0.002 & 0.003 & 0.20 & 0.21 & 9.7 & 24.9 \\
    \gem & 0.70 & 0.68 & 0.014 & 0.017 & 0.25 & 0.28 & 0.1 & 0.1 \\
    \merf & 0.66 & 0.67 & 0.024 & 0.018 & 0.52 & 0.51 & 0.1 & 0.1 \\
    \ddpm & 0.41 & 0.40 & 0.063 & 0.068 & 0.76 & 0.80 & 6.7 & 4.1 \\
    \bottomrule
    \end{tabular}
    }
\end{table}
}

\vspace{0.5mm}
\noindent \Rtwo{\textbf{Over-preprocessing Problem.} In both \cref{rare value merge} and \cref{PrivTree}, we apply a threshold $\beta$ to filter out those attributes with small domains to avoid ``over-preprocessing". In \Cref{enforce pre}, we present the synthesis results for ACSincome with mandatory preprocessing, which does not need preprocessing under threshold $\beta$. We can observe that preprocessing does not significantly influence the algorithms' utility. We believe that, because these datasets are inherently simple, preprocessing does not overly distort their information and thus has little impact on the synthesis results. A notable finding is that \rapp and \gsd run faster on the preprocessed dataset. We hypothesize that this is because preprocessing combines similar or less informative values together, which allows these two methods to capture the data characteristics more easily and converge more quickly. In summary, preprocessing can be unnecessary for those simple datasets, but mandatory preprocessing is likely not to be detrimental to synthesis performance.}

\vspace{0.5mm}
\noindent \textbf{Numerical Discretization Ablation.} Firstly, we summarize the domain sizes under different discretization results in \cref{bin num}. It is obvious that PrivTree always generates significantly fewer bins to represent numerical attributes, even for attributes with highly complex value distributions. This approach not only reduces noise in feature measurements but also simplifies the features, raising concerns about losing information. To better demonstrate the influence of discretization, we conduct the ablation study on Higgs-small dataset, which contains the most complex numerical variables among all datasets. For this ablation, we focus on $\varepsilon=1.0$, as is standard practice~\cite{du2024towards, tao2022benchmarkingdifferentiallyprivatesynthetic}. The results are shown in \cref{discretizer result}.

{
\setlength{\abovecaptionskip}{3pt}
\setlength{\belowcaptionskip}{-2pt}
\begin{table}[t]
    \centering
    \caption{Marginal-based methods' performance on the Higgs-small with different discretization under $\varepsilon = 1.0$. Here, ``Tree" means PrivTree and ``Uniform" means uniform binning.}
    \label{discretizer result}
    \setlength{\tabcolsep}{3.6pt}
    \resizebox{0.6\columnwidth}{!}{
        \begin{tabular}{l|cc|cc|cc|cc}
        \toprule 
            \multirow{2}{*}{\textbf{Method}} & \multicolumn{2}{c|}{\textbf{ML Eff. $\uparrow$}} & \multicolumn{2}{c|}{\textbf{Query Err. $\downarrow$}} & \multicolumn{2}{c|}{\textbf{Fid Err. $\downarrow$}} & \multicolumn{2}{c}{\textbf{Time (min) $\downarrow$}}\\ 
            \cline{2-9}
            & \rule{0pt}{1.1em} Tree & Uniform  & Tree & Uniform  & Tree & Uniform & Tree & Uniform  \\ 
        \midrule
            \privsyn & $0.43$ & $0.43$ & $0.005$ & $0.004$ &$0.19$ & $0.20$ &$2.1$ & $6.0$\\
            \privmrf & $0.65$ & $0.64$ & $0.005$ & $0.003$ &$0.19$ & $0.16$ &$14.7$ & $7.1$ \\
            \aim     & $0.67$ & $0.65$ & $0.005$ & $0.003$ &$0.20$ & $0.19$ &$689.6$ & $18.5$  \\
            \rapp    & $0.55$ & $0.53$ & $0.030$ & $0.029$ & $0.55$ & $0.65$ &$74.6$ & $40.0$  \\
            \gsd     & $0.50$ & $0.47$ & $0.043$ & $0.044$ &$0.57$ & $0.65$ &$58.2$ & $58.3$ \\
            \gem     & $0.56$ & $0.54$ & $0.019$ & $0.061$ &$0.29$ & $0.55$ &$0.4$ & $6.1$ \\   
        \bottomrule
        \end{tabular}
    }
    \vspace{-4mm}
\end{table}
}

{
\setlength{\abovecaptionskip}{3pt}
\begin{figure}[t]
    \centering
    \includegraphics[width=0.55\columnwidth]{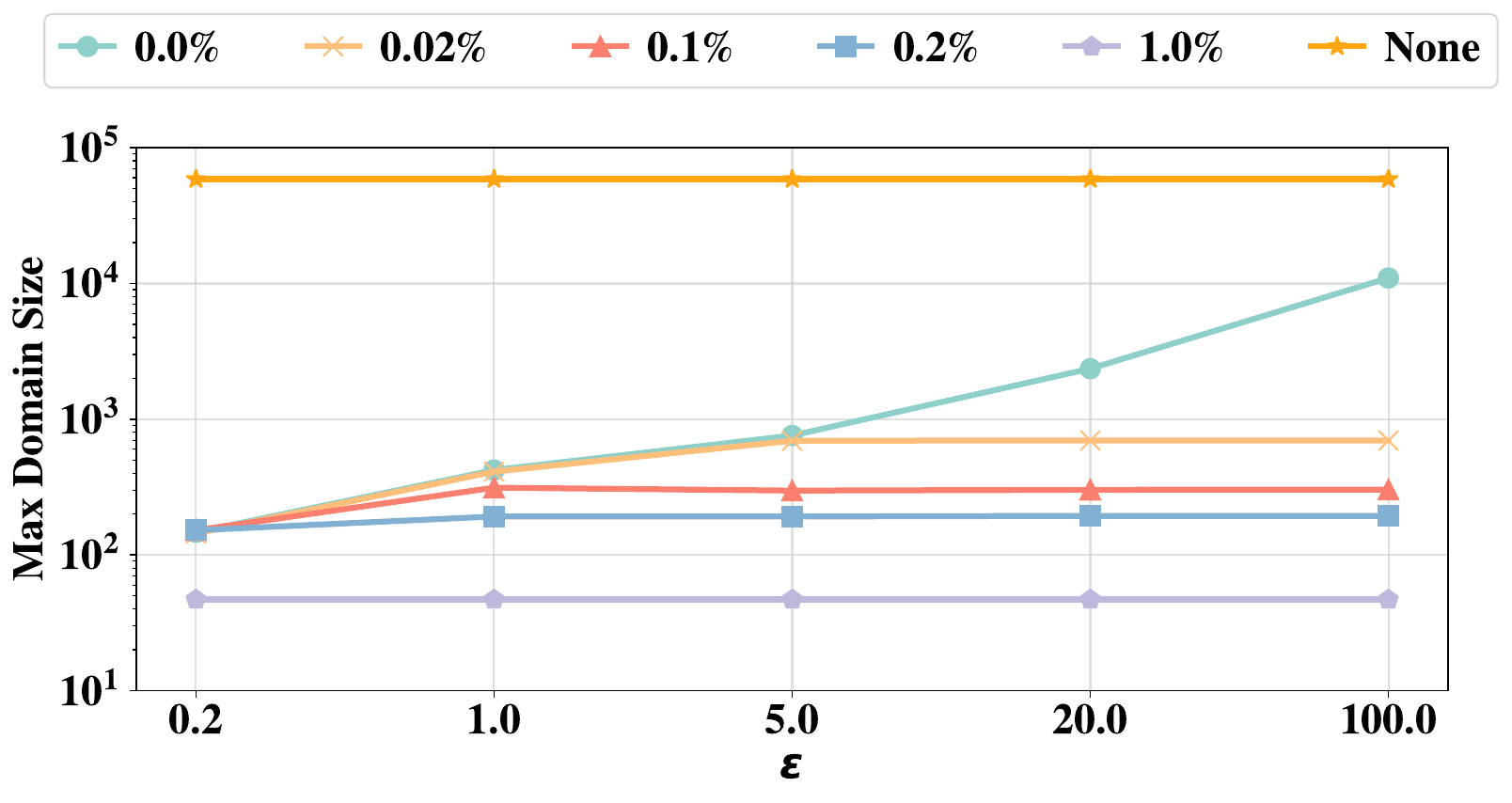}
    \caption{The maximum attribute domain size of Loan dataset under fixed merging thresholds and different $\varepsilon$. In the figure, ``$0.0\%$'' means no fixed threshold (merge only by $3\sigma$ criteria), and ``None'' means raw data.}
    \label{tab: exp merge}
\end{figure}
}

{
\setlength{\abovecaptionskip}{3pt}
\setlength{\belowcaptionskip}{3pt}
\begin{table}[t]
\footnotesize
    \centering
    \caption{Preprocessed numerical attributes' domain sizes under different discretization algorithms. The result is obtained on Higgs-small dataset and under setting $\varepsilon = 1.0$.}
    \label{bin num}
    \resizebox{0.55\columnwidth}{!}{
    \begin{tabular}{l|ccc|ccc}
    % {\linewidth}{l|>{\centering\arraybackslash}X>{\centering\arraybackslash}X>{\centering\arraybackslash}X}
    \toprule 
        \multirow{2}{*}{\textbf{Dataset}} & \multicolumn{3}{c|}{\textbf{Min Domain Size}} & \multicolumn{3}{c}{\textbf{Max Domain Size}} \\ \cline{2-7}
         & \rule{0pt}{1.1em} Raw &  PrivTree  & Uniform & Raw &  PrivTree  & Uniform \\ 
        \hline
        \rule{0pt}{1.1em}Bank & $505$ & $25$ & $100$ & $6024$ & $28$ & $100$\\
        Higgs-small & $4870$ & $6$ & $100$ & $60696$ & $18$ & $100$ \\
        Loan    & $101$ & $9$ & $100$ & $93995$ & $32$ & $100$\\
        \bottomrule 
    \end{tabular}}
\end{table}
}

In most cases, PrivTree performs better on deep learning tasks but is slightly worse in query errors and fidelity errors compared with uniform binning. We believe fewer bins help capture overall correlations by introducing less noise, which benefits deep learning tasks. However, query tasks and fidelity are more sensitive to exact values, and more bins lead to higher accuracy. Two exceptions are \gsd and \gem, where PrivTree outperforms uniform binning. We hypothesize that this is because PrivTree reduces the dimensionality of variables, thereby improving the convergence. These results do not completely align with observations in Tao et al.'s work~\cite{tao2022benchmarkingdifferentiallyprivatesynthetic}. We would propose three possible reasons. Firstly, Higgs-small dataset is more value-rich than the datasets used in their work, which can lead to high sensitivity to binning methods. Secondly, the evaluation metrics are different, while ours focus more on measurements on higher-dimensional marginals (e.g., 3-way marginals). This can influence the results. Finally, the investigated algorithms vary in the two works, which may lead to different comparison conclusions.  Another unusual finding is that under PrivTree binning, \aim shows a significantly worse runtime. We believe this is because smaller domain sizes will cause \aim to allocate budget to more marginals. This may lead to large cliques, slowing down the execution speed of PGM.

We can conclude that different discretization methods have their own advantages and weaknesses. It is important to choose an appropriate method for different algorithms. For example, for those methods based on generative networks, PrivTree could be a potentially better preprocessing method because it can significantly decrease the dimension of models by reducing domain complexity. For \mtdtt{PGM} and \mtdtt{GUM}, uniform binning could be a better choice due to their stronger ability to fit marginals with large domains.

\vspace{0.5mm}
\noindent \textbf{Category Merging Ablation.} Finally, we briefly discuss the necessity of introducing a fixed merging threshold. We compare the maximum domain size of preprocessed categorical attributes under different fixed merging thresholds in \cref{tab: exp merge}. We can conclude that domain size cannot be reduced efficiently under large $\varepsilon$ if we do not apply a fixed merging threshold ($0.0\%$). Furthermore, we may over-merge categories if this threshold is large (e.g., $1\%$), which may cause potential synthesis errors.

%---------------------------------------------------------------------
%---------------------------------------------------------------------

{
\setlength{\abovecaptionskip}{2pt}
\setlength{\belowcaptionskip}{0pt}
\begin{figure*}[t]
    \centering
    \begin{subfigure}[t]{\textwidth}
        \centering
        \includegraphics[width=\textwidth]{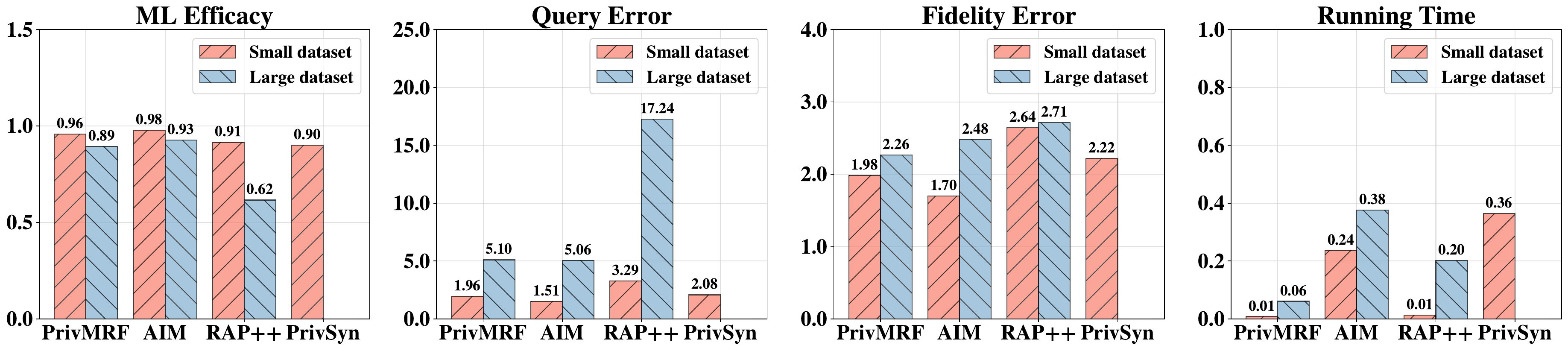}
    \end{subfigure}
    \vspace{2mm}
    \begin{subfigure}[t]{\textwidth}
        \centering
        \includegraphics[width=\textwidth]{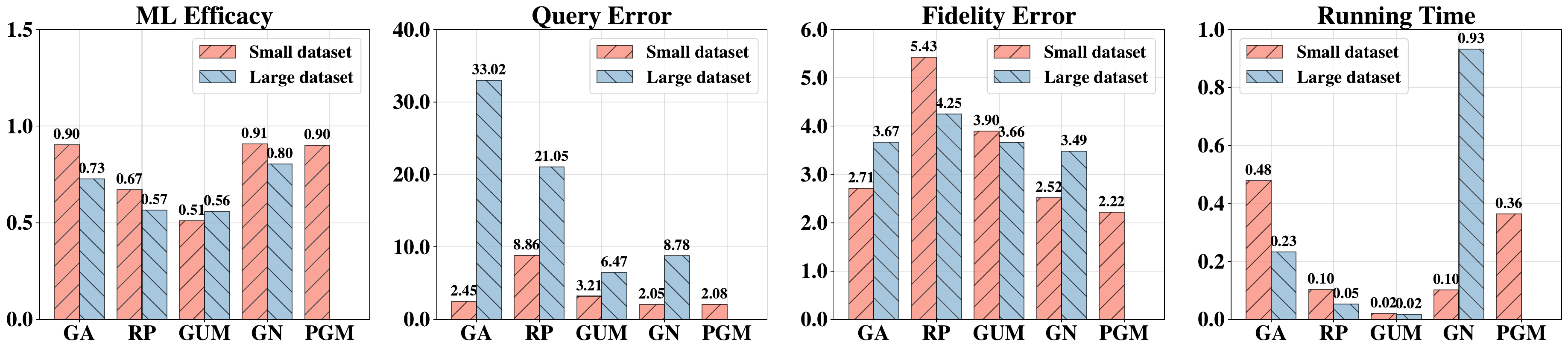}
    \end{subfigure}
    \caption{Average scaled evaluations for different marginal selection modules (first row) and synthesis modules (second row) on different datasets. We use PGM as the common data synthesizer and \privsyn as the common marginal selector. The running time is scaled by min-max linear normalization, and others are scaled using ground truth values provided in \cref{general result}. The small datasets include ACSincome and ACSemploy, whereas the large datasets consist of Bank, Higgs-small, and Loan.}
    \label{fig3: exp}
\end{figure*}
}

\subsection{Module Comparison}
\label{divide exp}

In this subsection, we decompose our experiments into two parts: one focusing on different feature selection algorithms and the other on different synthesis modules. In other words, we fix either the selection or the synthesis approach and then evaluate how different algorithms perform under that fixed condition. Since some features are not compatible with other synthesis methods (e.g., random Fourier features), we only focus on marginal-based methods, namely \privsyn, \rapp, \gsd, \aim, \privmrf, and \gem.

To make clearer comparisons, we divide the datasets into two groups. The first group, ACSincome and ACSemploy, is relatively low-dimensional with moderately sized attributes' domains. We call them ``small datasets'' for description simplicity. The second group, Bank, Higgs-small, and Loan, is higher-dimensional or has larger attribute domains, which are called ``large datasets''. To make a better comparison, we scale each metric and display the average performance on these two groups of datasets under different $\varepsilon$.

\vspace{0.5mm}
\noindent \textbf{Comparison of Feature Selection Modules.} Since some algorithms use the same or similar marginal selection methods, or have not specified them, we consider four marginal selection methods included in previous work. They are: \privsyn selection, \privmrf selection, \rapp selection, and \aim selection. The synthesis processes for these four selection methods are all set to \mtdtt{PGM}. \cref{fig3: exp} show the average scaled performances.

Among the selection methods, \privsyn and \rapp show weaker fitting utility, consistent with our analysis. \privsyn fails to use intermediate information during selection, reducing its ability to capture necessary data features and leading to excessive selections. This also overwhelms \mtdtt{PGM} on larger datasets due to high memory demands. Similarly, \rapp employs the Gumbel mechanism to select multiple marginals per iteration, without fully accounting for intermediate results. It also decreases the number of synthesis rounds required, thereby accelerating the algorithm's execution. Moreover, its selection criterion focuses solely on marginal query error, ignoring noise error, which is unbalanced and potentially influences the algorithm's performance.

The utility performances of \aim and \privmrf, are very similar since both perform a single marginal refinement after updating the intermediate information. A notable observation is that \privmrf has a shorter running time than \aim. We attribute this to the well-designed initial marginal set in \privmrf, which decreases the need for further refinement and speeds up the total algorithm.

%----------------------------------------------------------------
%----------------------------------------------------------------

\vspace{0.5mm}
\noindent \textbf{Comparison of Synthesis Modules.} We compare the synthesis algorithms of all six algorithms: \mtdtt{GUM}, \mtdtt{PGM}, relaxed projection (RP), genetic algorithm (GA), and generative network (GN). \mtdtt{PGM} is employed by both \aim and \privmrf. To avoid multiple fitting steps from adaptive selection and control the running time, we use the \privsyn selection algorithm as the common marginal selector. The results are shown in \cref{fig3: exp}. \mtdtt{GUM} and \mtdtt{PGM} are conducted on CPU, and the remaining works are on GPU.

Among these methods, \mtdtt{PGM} and the generative network achieve the best overall accuracy and stability in fitting features on small datasets. \mtdtt{PGM} benefits from its expressive structure, which, as indicated by \cref{eq:pgm}, infers marginals without refitting existing ones, thus minimizing compounding errors. However, the PGM's fitting process is quite slow, due to the densely selected marginals by \privsyn, causing it to fail to deal with large datasets. Indeed, PGM is most efficient when coupled with a marginal selection procedure that is designed to make the resulting graphical model ``tree-like'' \cite{mckenna2021winning,cai2021data,mckenna2022aim}.  The generative network’s strong representational capacity also supports accurate feature fitting on small datasets. However, when it turns to larger and more complex datasets, its superiority in efficiency and utility will be diminished due to the greater difficulty of training high-dimensional models.

In contrast, \mtdtt{GUM}, the genetic algorithm, and relaxed projection demonstrate weaker fitting capabilities. \mtdtt{GUM} refines marginals individually, overlooking global correlations. However, it shows a superiority in running time, aligning with our analysis. The genetic algorithm’s reliance on randomness makes it unsuitable for complex, high-dimensional datasets with large attribute domains under time constraints, leading to poor performance on large datasets. Relaxed projection suffers similarly. In high-dimensional spaces with extensive attributes' domains, the encoded data dimension becomes excessively large, complicating optimization and making it harder to converge to the optimal point, and finally resulting in poor performance.

% \begin{keytakeaway}[frametitle={Key Takeaways}]
%     The selection strategies employed by \aim and \privmrf are relatively optimal due to their efficient usage of intermediate measurements and reasonable initialization. Moreover, the performance of various synthesis methods varies significantly, and all existing synthesis methods exhibit certain weaknesses. 
% \end{keytakeaway}

\section{Conclusion and Discussion}
\label{sec: conclusion}

Data synthesis under differential privacy remains a critical and active area of research. Despite a growing number of methods being proposed, the field still lacks a fair and comprehensive comparative analysis. In this paper, we conduct analysis, comparison, and evaluation of several methods on a unified framework. The primary findings of our study are as follows: 
\begin{enumerate}[label=\textbullet, leftmargin=*]
    \item \emph{A significant trade-off exists between utility and efficiency, resulting in no single algorithm dominating the others.} deep learning-based methods, though highly efficient, generally exhibit inferior performance compared to traditional statistical methods. \Rtwo{Nonetheless, under current implementations, some statistical methods, such as \aim and \privmrf, are often time-consuming, which is a trade-off for superior utility.}
    \item \emph{Data preprocessing strategy is important but algorithm-dependent.} Data preprocessing is crucial in reducing algorithm complexity without introducing too much error. The choice of data preprocessing methods should align with the working principles of the algorithm. Appropriate preprocessing techniques can enhance both the efficiency and effectiveness of the algorithms, whereas unsuitable methods may hinder their performance. 
    \item \emph{Improvement on data synthesis module remains a promising direction.} We find that some feature selection methods outperform others, but existing synthesis modules exhibit significant drawbacks in different ways. For example, even though PGM demonstrates outstanding utility, it cannot work on densely selected marginals (due to high memory requests and long execution time), while generative networks face significant challenges when dealing with high-dimensional data. There remains substantial room for improvement in developing algorithms that can efficiently, reliably, and accurately generate data. 
\end{enumerate}

{
\setlength{\abovecaptionskip}{4pt}
\setlength{\belowcaptionskip}{2pt}
\begin{table}[t]
\footnotesize
    \centering
    \caption{\Rthree{Estimated cost (US dollar) of running different algorithms.}}
    \vspace{-1mm}
    \label{price table}
    \resizebox{0.98\columnwidth}{!}{
    \begin{tabular}{>{\centering\arraybackslash}p{0.1\columnwidth} *{8}{|>{\centering\arraybackslash}p{0.11\columnwidth}}} 
    \toprule 
        \textbf{Method} & \textbf{\privsyn} & \textbf{\privmrf} & \textbf{\aim} & \textbf{\rapp} & \textbf{\gsd} & \textbf{\gem} & \textbf{\merf} & \textbf{\ddpm}\\
        \midrule
        \textbf{Total Cost} & \$ 0.2 & \$ 1.1 & \$ 5.7 & \$ 68.1& \$ 17.0 & \$ 0.7 & \$ 0.1 & \$ 2.1 \\
    \bottomrule 
    \end{tabular}
    }
\end{table}
}

Our work not only sheds light on the strengths and limitations of current methods but also identifies several promising directions for future research. \Rtwo{These include tailoring preprocessing strategies to algorithmic needs, developing more effective feature selection techniques, and improving the utility of deep learning models or the efficiency of some statistical methods by more efficient implementation or GPU acceleration.} 
\Rthree{
In addition to these research opportunities, there are still certain other aspects, though not investigated in this work, worth exploring. For example, beyond the time efficiency perspective, dollar cost is a more realistic factor in algorithm selection. Based on our experimental setups and hardware, we roughly set the cost at \$0.8/hr for GPU usage, and \$0.2/hr for CPU usage (the actual prices may vary due to many factors such as dataset complexity, which can affect our selection of hardware, and price fluctuations of cloud service~\cite{price1,price2,price3,price4}, etc.). We summarize the total price consumption of all algorithms during the overall comparison in \Cref{price table}. We can observe that those methods with lower time efficiency tend to incur higher costs.
}

\Rthree{
Finally, we discuss the scalability of our framework. It offers a broader perspective for analyzing existing synthesis algorithms, but we also need to recognize its limitations. For example, we regard gradients as features to explain those methods based on DP-SGD. But the gradient feature itself is not interpretable enough. Besides, some ``model-level DP" methods may fall outside our three-step workflow. For instance, PATE-GAN~\cite{jordon2018pate} uses multiple teacher discriminators to achieve DP and trains the student discriminator on them to further guide the data generator. In this case, our framework may lack corresponding explanatory power and require further extension to explain these methods.
}

\begin{acks}
We thank anonymous reviewers for their valuable comments. This paper was partially supported by NSF CNS-2220433, NSF OAC-2319988, and a CCI award.
\end{acks}

\bibliographystyle{ACM-Reference-Format}
\bibliography{ref}

%%
%% If your work has an appendix, this is the place to put it.
\appendix

\section{DP Preliminaries}

\subsection{RDP Properties}

We briefly introduce RDP composition and post-processing properties in \cref{theorem: compos} and \cref{theorem: post-p}.
\begin{theorem}[Composition] \label{theorem: compos}
    Let $f: \mathcal{D} \rightarrow \mathcal{R}_1$ be $(\alpha, \varepsilon_1)$-RDP and $g: \mathcal{R}_1 \times \mathcal{D} \rightarrow \mathcal{R}_2$ be $(\alpha, \varepsilon_2)$-RDP respectively. Then the mechanism defined as $(X,Y)$, where $X \sim f(D)$ and $Y \sim (D, f(D))$, satisfies $(\alpha, \varepsilon_1 + \varepsilon_2)$-RDP.
\end{theorem}

\begin{theorem}[Post-Processing] \label{theorem: post-p}
     Let $f: \mathcal{D} \rightarrow \mathcal{R}_1$ is $(\alpha, \varepsilon)$-RDP, and $g: \mathcal{R}_1 \rightarrow \mathcal{R}_2$ is an arbitrary randomized mapping. Then $g\circ f$ is also $(\alpha, \varepsilon)$-RDP.
\end{theorem}

\subsection{DP Mechanism}
Before introducing DP mechanisms, a key quantity needed to be defined is the sensitivity: 
% \gcc{Dose this sentence correct? It odds to me.}.
\begin{definition}[Sensitivity]
    Let $f: \mathcal{D} \rightarrow \mathcal{R}^k$ be a vector-valued function of the input data, then the $\ell_2$ sensitivity of $f$ is defined as 
    \[
        \Delta_f = \max_{D \simeq D'} \left\lVert f(D) - f(D') \right\rVert_2
    \]
\end{definition}

\noindent \textbf{Gaussian Mechanism}. Gaussian Mechanism (GM)~\cite{mironov2017renyi}, which adds noise sampled from Gaussian distribution, has widely been used to achieve $(\alpha, \varepsilon)$-RDP. 
Specifically, Let $f$ be a vector-valued function of the input data. The Gaussian Mechanism adds i.i.d. Gaussian noise with scale $\sigma \Delta_f$ to each entry of $f$.
\begin{equation}\label{eq:gaussian}
    \mathcal{A}(D) = f(D) + \sigma \Delta_f \mathcal{N}\left(0, \; \mathbb{I}\right),
\end{equation}
where $\mathcal{N}$ refers to Gaussian distribution. The RDP guarantee of GM is given by the \cref{theorem:gm}. 
\begin{theorem} \label{theorem:gm}
    The Gaussian Mechanism defined above satisfies $\left(\alpha, \frac{\alpha}{2 \sigma^2}\right)$-RDP.
\end{theorem}

% \begin{figure}[t]
%     \centering
%     \includegraphics[width=\columnwidth]{fig/intro.pdf}
%     \caption{An illustration of how DP data synthesis works and protects the privacy of data against potential adversaries.}
%     \label{fig:intro}
%     \vspace{-5mm}
% \end{figure}

\noindent \textbf{Exponential Mechanism}. Let $q_r$ be a score function for all $r \in \mathcal{R}$. Then the exponential mechanism (EM)~\cite{mckenna2022aim, liu2021iterative} outputs a candidate $r$ according to the following distribution:
\begin{equation}
    \Pr[\mathcal{A}(D) = r] \propto \exp\left(\frac{\epsilon}{2\Delta}\cdot q_r(D)\right),
\end{equation}
where $\Delta = \max_{r \in \mathcal{R}} \Delta(q_r)$. The RDP guarantee of EM is provided by \cref{theorem:em}
\begin{theorem} \label{theorem:em}
    The Exponential Mechanism defined above satisfies $\left(\alpha, \frac{\alpha \epsilon^2}{8}\right)$-RDP for $\forall \alpha > 1$.
\end{theorem}

\noindent \textbf{Gumbel Noise}. Reporting noise max with Gumbel noise is a derivative mechanism from the exponential mechanism used for marginal query selection by some previous work~\cite{vietri2020neworacleefficientalgorithmsprivate, vietri2022private}. Specifically, it outputs $i^* = \arg \max_{i \in [m]} \{|q_i(D) - a_i| + Z_i\}$, where $q_i$ is the marginal query function defined in ~\cite{vietri2020neworacleefficientalgorithmsprivate, vietri2022private}; $a_i$ is the query answer and $Z_i \sim \text{Gumbel}(1/n\sqrt{2\rho})$. This mechanism has been proven to satisfy $\left(\alpha, \alpha\rho\right)$-RDP for $\forall \alpha > 1$.

\vspace{1.1mm}
\noindent \textbf{DP-SGD}. Differentially private stochastic gradient descent (DP-SGD) is the most popular way to train the model to satisfy DP. In the training process, we assume that $\mathcal{L}$ is the loss function, and we have a clipping function defined by $\text{Clip}_C(g) = \min\left\{1, \frac{C}{\lVert g \rVert_2}\right\} g$ and a Gaussian noise level $\sigma$~\cite{mironov2019r}. The DP-SGD can be expressed as 
$$
    \theta \leftarrow \theta - \eta \left( \frac{1}{|b|} \sum_{i \in b} \text{Clip}_C(\nabla\mathcal{L}(\theta, x_i)) + C \mathcal{N}(0, \sigma^2 \mathbf{I}) \right) 
$$
Here $\eta$ is the learning rate, $\nabla\mathcal{L}(\theta, x_i)$ is the gradient of the loss function $\mathcal{L}$ in relation to model parameters $\theta$ and data point $x_i$ in sample batch $b$. By clipping the gradient, we control the sensitivity and thus can apply the Gaussian mechanism to guarantee DP.

\section{DP Guarantee of Preprocessing Algorithms}
\label{appendix: algo}

\subsection{PrivTree Binning}
The privacy guarantee is given by the following lemma.
\begin{lemma}
\label{privtree dp}
    For any $\alpha > 1$, \cref{PrivTree} satisfy $(\alpha, \alpha \rho_1 )$-R\'enyi differential privacy.
\end{lemma}

This proof is organized by two steps: firstly we proof that for each attribute, Algorithm\cref{PrivTree} can achieve a $(\alpha, \alpha \rho_1/K)$-R\'enyi DP, then by composition theorem, we can draw the conclusion that Algorithm\cref{PrivTree} can achieve $(\alpha, \alpha \rho_1)$-R\'enyi DP totally. 

By the privacy proof of PrivTree in Zhang et al.'s work~\cite{zhang2016privtree}, we have that if 
\[
    \lambda' \ge \frac{2\beta-1}{\beta-1} \cdot \frac{1}{\varepsilon} 
\]
and 
\[
    \delta' = \lambda' \cdot \ln{\beta} 
\]
single PrivTree algorithm satisfies $\varepsilon$-DP. Replacing $\varepsilon$ and $\beta$ by the definition in Algorithm\cref{PrivTree}, we have that each round of PrivTree binning satisfies $\frac{2\rho_1}{|V_n|}$-DP

Then referring to Bun et al.'s work~\cite{bun2016concentrated}, we have that if a mechanism satisfies $\varepsilon$-DP, it also satisfies $(\alpha, \frac{1}{2}\varepsilon^2 \alpha$-RDP. Therefore we obtain that PrivTree is $(\alpha, \frac{\rho_1}{|V_n|}\alpha)$-RDP. Because we need to conduct PrivTree binning for $|V_n|$ rounds, by composition theorem of RDP, we have that Algorithm\cref{PrivTree} satisfies $(\alpha, \rho_1\alpha)$-RDP.

\subsection{Rare Category Merging}
For each attribute that needs preprocessing, we equally divide the privacy budget and use it to determine those categories whose frequency is lower than the threshold. These categories will be replaced by the same encoding category. The privacy guarantee can be formalized in the following lemma. 
\begin{lemma}
\label{rare dp}
    For any $\alpha > 1$, \cref{rare value merge} satisfies $(\alpha, \alpha \rho_2 )$-R\'enyi differential privacy.
\end{lemma}

The proof of this lemma can be easily obtained by the property of the Gaussian mechanism. Referring to \cref{theorem:gm}, we know that adding Gaussian noise with $\sigma = \sqrt{\frac{1}{2\rho'}}$ satisfying $(\alpha, \alpha \rho' )$-R\'enyi differential privacy for any $\alpha > 1$. Then by combining the fact that $\rho' = \rho_2 / |V_c|$ and the composition property of RDP, we have that the total algorithm satisfies $(\alpha, \alpha \rho_2)$-R\'enyi differential privacy for any $\alpha > 1$.

\section{Missing Proof} \label{appendix: proof}

In this section, we provide the proof of ~\cref{theorem: adapt measure}. Before we prove this theorem, we give a lemma as follows.
\begin{lemma}\label{lemma kl}
    Assuming that 
    \[
    P_1(z) = \sum_x p(x)P_1(z|x) \;\;\; \text{and} \;\;\; P_2(z) = \sum_x p(x)P_2(z|x),
    \]
    we have 
    \[
    \dkl\left(P_1(z) \;\|\; P_2(z)\right) \leq \sum_x p(x) \dkl\left(P_1(z|x) \;\|\; P_2(z|x)\right)
    \]
\end{lemma}
\noindent Firstly, we have 
\[
\ln\left( \frac{P_1(z)}{P_2(z)}\right) = \ln\left( \frac{\sum_x p(x)P_1(z|x)}{\sum_x p(x)P_2(z|x)} \right).
\]
We already have ``log-sum'' inequality~\cite{csiszar1967information}, which can be expressed as 
\[
\sum_{i=1}^n x_i \log\left(\frac{x_i}{y_i}\right)
\;\ge\;
\left(\sum_{i=1}^n x_i\right)\,\log\left(\frac{\sum_{i=1}^n x_i}{\sum_{i=1}^n y_i}\right).
\]
Let $\alpha_x(z) = \frac{p(x)P_1(z|x)}{P_1(z)}$, by ``log-sum'' inequality, we have
\[
\begin{aligned}
\ln\left( \frac{P_1(z)}{P_2(z)}\right) & \leq \sum_x \alpha_x(z) \ln\left( \frac{ p(x)P_1(z|x)}{ p(x)P_2(z|x)} \right) \\
& = \sum_x \alpha_x(z) \ln\left( \frac{P_1(z|x)}{P_2(z|x)} \right)
\end{aligned}
\]
Taking mathematical expectations, we have 
\[
\begin{aligned}
\mathbb{E}_{z\sim P_1}\left[ \ln\left( \frac{P_1(z)}{P_2(z)}\right) \right] \leq \mathbb{E}_{z\sim P_1}\left[ \sum_x \alpha_x(z) \ln\left( \frac{P_1(z|x)}{P_2(z|x)} \right) \right]
\end{aligned}
\]
Notice that $\alpha_x(z) = P(x|z)$, we can rewrite the right-hand side of the above inequality as 
\[
\begin{aligned}
&\sum_{x}\mathbb{E}_{x,z\sim p(x), P_1(z|x)}\left[ \mathbb{I}(X=x) \ln\left( \frac{P_1(z|x)}{P_2(z|x)} \right) \right] \\ 
= & \sum_x p(x) \dkl\left(P_1(z|x) \;\|\; P_2(z|x)\right)
\end{aligned}
\]
Combining all above, we have proved that 
\[
\dkl\left(P_1(z) \;\|\; P_2(z)\right) \leq \sum_x p(x) \dkl\left(P_1(z|x) \;\|\; P_2(z|x)\right).
\]

Now we give the proof of \cref{theorem: adapt measure}. For the left-hand side of Equation~(\ref{eq: theorem: adapt measure}), we have 
\begin{equation} \label{eq: lhs adapt measure}
    \dkl\left(\Pr[A_i, A_j] \left\|\; \Pr[A_i]\Pr[A_j] \right.\right) = I\left(A_i, A_j\right),
\end{equation}
where $I$ refers to mutual information~\cite{shannon1948mathematical}. For the right-hand side of Equation~(\ref{eq: theorem: adapt measure}), applying ~\cref{lemma kl}, we have the following property: 
\begin{equation} \label{eq: rhs adapt measure}
\begin{aligned}
    & \dkl\left(\left. \Pr[A_i, A_j] \;\right\| \right. \\
    & \; \sum_{A_1, \cdots, A_k} \Pr[A_1, \cdots, A_k] \cdot \Pr[A_i | A_1, \cdots, A_k] \Pr[A_j | A_1, \cdots, A_k] ) \\
    = & \dkl\left(\left.\sum_{A_1, \cdots, A_k} \Pr[A_1, \cdots, A_k] \Pr[A_i, A_j|A_1, \cdots, A_k] \;\right\| \right.\\
    & \; \left.\sum_{A_1, \cdots, A_k} \Pr[A_1, \cdots, A_k] \cdot \Pr[A_i | A_1, \cdots, A_k] \Pr[A_j | A_1, \cdots, A_k]\right) \\
    \leq & \sum_{A_1, \cdots, A_k} \Pr[A_1, \cdots, A_k] \cdot \dkl\left.\left(\Pr[A_i, A_j|A_1, \cdots, A_k] \right. \;\right\| \\
    &\quad \quad \quad \quad \quad \quad \quad \quad \quad \left. \Pr[A_i | A_1, \cdots, A_k] \Pr[A_j | A_1, \cdots, A_k]\right) \\
    = & I\left(A_i, A_j \;|\; A_1, \cdots, A_k\right) 
\end{aligned}
\end{equation}
% \xl{I checked the Jessen inequality, and I'm not sure whether the second $\sum M_{1, \cdots, k}$ should also be extracted.}
% \xl{I'm not sure the correctness of the final step in the proof.}
where $I(\cdot \;|\; \cdot)$ is the conditional mutual information. 
Then by property of mutual information~\cite{thomas2006elements}, we have 
\begin{equation} \label{eq: mutual info}
I\left(A_i, A_j \;|\; A_1, \cdots, A_k\right) \leq I\left(A_i, A_j\right).
\end{equation}
Combining \cref{eq: lhs adapt measure}, \cref{eq: rhs adapt measure} and \cref{eq: mutual info}, we can prove \cref{theorem: adapt measure}.

\section{Missing Experimental Settings} \label{appendix: param}

\subsection{Supplementary Dataset information}
ACSincome and ACSemploy~\cite{ding2021retiring} are both drawn from 2018 national census data.  
The Bank dataset is on the UCI open dataset website~\cite{bank_marketing_222}, which is related to direct marketing campaigns of a Portuguese banking institution. 
Higgs-small~\cite{higgsdata} dataset was produced using Monte Carlo simulations, which include different features of particles in the accelerator. 
Loan dataset~\cite{loandata} was derived from data LendingClub issued through 2007-2014.

\subsection{Implementations}

We implement some of the algorithms referring to their open-source code bases. Here, we list them below. 
\begin{enumerate}[leftmargin=*, label=\textbullet]
    \item \privsyn: \url{https://github.com/agl-c/deid2_dpsyn}
    \item \privmrf: \url{https://github.com/caicre/PrivMRF}
    \item \rapp: \url{https://github.com/amazon-science/relaxed-adaptive-projection}
    \item \aim: \url{https://github.com/ryan112358/private-pgm}
    \item \gsd: \url{https://github.com/giusevtr/private_gsd}
    \item \gem: \url{https://github.com/terranceliu/iterative-dp?tab=readme-ov-file}
    \item \merf: \url{https://github.com/ParkLabML/DP-MERF}
    \item \ddpm: \url{https://github.com/yandex-research/tab-ddpm}
\end{enumerate}

\subsection{Hyperparameters for Preprocessing algorithms}

We apply uniform discretization and rare category merging for all algorithms as the default aligned preprocessing methods. Moreover, as shown in \cref{info:datasets}, some attributes only contain a few unique values, which are simple enough to handle without additional preprocessing. Applying preprocessing to such attributes is unnecessary and may introduce more errors. Therefore, we preprocess attributes only when their domain size exceeds 100. By default, the number of bins is set to 100, the fixed merging threshold $\theta$ to $0.2\%$, and the privacy budget proportion allocated to the preprocessing step to $10\%$.

\subsection{Hyperparameters for Full Algorithms}
We list the algorithms' hyperparameters in \cref{overall exp}. 
% \privsyn and \privmrf are not listed here, because we all use the default hyperparameters referring to the original works. 
From here on out, unless otherwise specified, INC refers to the ACSincome dataset; EMP refers to the ACSemploy dataset; BK refers to the Bank dataset; HIG refers to the Higgs-small dataset; LN refers to the Loan dataset. 

{
\setlength{\abovecaptionskip}{3pt}
\begin{table}[H]
\footnotesize
\centering
\caption{PrivSyn Hyperparameters}
% \resizebox{\columnwidth}{!}{
\begin{tabularx}{0.7\columnwidth}{l|XXXXX}
     \toprule
     \textbf{Hyperparameter} & \textbf{INC}& \textbf{EMP}& \textbf{BK}& \textbf{HIG} & \textbf{LN}\\
     \midrule
     Consistent iteration & 501& 501& 501& 501& 501\\
     Max update iteration & 50& 50& 50& 50& 50\\
     \bottomrule
\end{tabularx}
% }
\end{table}
}

{
\setlength{\abovecaptionskip}{3pt}
\begin{table}[H]
\footnotesize
\centering
\caption{AIM Hyperparameters}
% \resizebox{\columnwidth}{!}{
\begin{tabularx}{0.75\columnwidth}{l|XXXXX}
     \toprule
     \textbf{Hyperparameter} & \textbf{INC}& \textbf{EMP}& \textbf{BK}& \textbf{HIG} & \textbf{LN}\\
     \midrule
     Max model size & $100$ & $100$ & $100$ & $100$ & $100$ \\
     Max iteration & $1000$ & $1000$ & $1000$ & $1000$ & $1000$ \\
     Max marginal size & $2.5e+5$ & $2.5e+5$ & $2.5e+5$ & $2.5e+5$ & $2.5e+5$ \\
     \bottomrule
\end{tabularx}
% }
\end{table}
}

{
\setlength{\abovecaptionskip}{3pt}
\begin{table}[H]
\footnotesize
\centering
\caption{Private-GSD Hyperparameters}
% \resizebox{\columnwidth}{!}{
\begin{tabularx}{0.75\columnwidth}{l|XXXXX}
     \toprule
     \textbf{Hyperparameter} & \textbf{INC}& \textbf{EMP}& \textbf{BK}& \textbf{HIG} & \textbf{LN}\\
     \midrule
     Mutation number & $50$ & $50$ & $50$ & $50$ & $50$ \\
     Crossover number & $50$ & $50$ & $50$ & $50$ & $50$ \\
     Upsample number & $1e+5$ & $1e+5$ & $1e+5$ & $1e+5$ & $1e+5$ \\
     Genetic iteration & $1e+6$ & $1e+6$ & $1e+6$ & $1e+6$ & $1e+6$ \\
     \bottomrule
\end{tabularx}
% }
\end{table}
}

{
\setlength{\abovecaptionskip}{3pt}
\begin{table}[H]
\footnotesize
\centering
\caption{PrivMRF Hyperparameters}
% \resizebox{\columnwidth}{!}{
\begin{tabularx}{0.75\columnwidth}{l|XXXXX}
     \toprule
     Hyperparameter & INC& EMP& BK& HIG & LN\\ 
     \midrule
     Graph construction parameter $\theta$ & $6$& $6$& $6$& $6$& $6$ \\ 
     Sample size $k$ & $400$& $400$& $400$& $400$& $400$ \\
     Estimation iteration & $3000$& $3000$& $3000$& $3000$& $3000$ \\
     Size penalty & $1e-8$& $1e-8$& $1e-8$& $1e-8$& $1e-8$ \\
     Max marginal attributes number & $6$& $6$& $6$& $6$& $6$ \\
     Max clique size & $1e+7$& $1e+7$& $1e+7$& $1e+7$& $1e+7$ \\
     \bottomrule
\end{tabularx}
% }
\end{table}
}

{
\setlength{\abovecaptionskip}{3pt}
\begin{table}[H]
\footnotesize
\centering
\caption{GEM Hyperparameters}
% \resizebox{\columnwidth}{!}{
\begin{tabularx}{0.75\columnwidth}{l|XXXXX}
     \toprule
     \textbf{Hyperparameter} & \textbf{INC}& \textbf{EMP}& \textbf{BK}& \textbf{HIG} & \textbf{LN}\\
     \midrule
     Synthesis size & $1024$ & $1024$ & $1024$ & $1024$ & $1024$ \\
     Learning rate & $1e-3$ & $1e-3$ & $1e-3$ & $1e-3$ & $1e-3$ \\
     Max iteration & $500$ & $500$ & $500$ & $500$ & $500$ \\
     Max selection round & $50$ & $85$ & $80$ & $140$ & $210$ \\
     \bottomrule
\end{tabularx}
% }
\end{table}
}

{
\setlength{\abovecaptionskip}{3pt}
\begin{table}[H]
\footnotesize
\centering
\caption{RAP++ Hyperparameters}
% \resizebox{\columnwidth}{!}{
\begin{tabularx}{0.75\columnwidth}{l|XXXXX}
     \toprule
     \textbf{Hyperparameter} & \textbf{INC}& \textbf{EMP}& \textbf{BK}& \textbf{HIG} & \textbf{LN}\\ 
     \midrule
     Random projection number & $2e+6$ & $2e+6$ & $2e+6$ & $2e+6$ & $2e+6$ \\
     Categorical optimization rate & $3e-3$ & $3e-3$ & $3e-3$ & $3e-3$ & $3e-3$ \\
     Numerical optimization rate & $6e-3$ & $6e-3$ & $6e-3$ & $6e-3$ & $6e-3$ \\
     Top q & $5$ & $5$ & $5$ & $5$ & $5$ \\
     Categorical optimization step & $1$ & $1$ & $1$ & $1$ & $1$ \\
     Numerical optimization step & $3$ & $3$ & $3$ & $3$ & $3$ \\
     Upsample rate & $10$ & $10$ & $20$ & $20$ & $40$ \\
     \bottomrule
\end{tabularx}
% }
\end{table}
}

{
\setlength{\abovecaptionskip}{3pt}
\begin{table}[H]
\footnotesize
\centering
\caption{DP-MERF Hyperparameters}
% \resizebox{\columnwidth}{!}{
\begin{tabularx}{0.75\columnwidth}{l|XXXXX}
     \toprule
     \textbf{Hyperparameter} & \textbf{INC}& \textbf{EMP}& \textbf{BK}& \textbf{HIG} & \textbf{LN}\\
     \midrule
     Random feature dimension & $2e+3$ & $2e+3$ & $2e+3$ & $2e+3$ & $2e+3$ \\
     Mini batch rate & $5e-2$ & $5e-2$ & $5e-2$ & $5e-2$ & $5e-2$ \\
     Epoch number & $1e+3$ & $1e+3$ & $1e+3$ & $1e+3$ & $1e+3$ \\
     Learning rate & $1e-2$ & $1e-2$ & $1e-2$ & $1e-2$ & $1e-2$ \\
     \bottomrule
\end{tabularx}
% }
\end{table}
}

{
\setlength{\abovecaptionskip}{3pt}
\begin{table}[!h]
\footnotesize
\centering
\caption{TabDDPM Hyperparameters}
% \resizebox{\columnwidth}{!}{
\begin{tabularx}{0.75\columnwidth}{l|XXXXX}
     \toprule
     \textbf{Hyperparameter} & \textbf{INC}& \textbf{EMP}& \textbf{BK}& \textbf{HIG} & \textbf{LN}\\
     \midrule
     Denoiser layer dimension & $256$ & $256$ & $256$ & $256$ & $1024$ \\
     Denoiser layer number & $2$ & $2$ & $2$ & $2$ & $2$ \\
     Epoch number & $50$ & $100$ & $100$ & $100$ & $100$ \\
     Batch size & $512$ & $512$ & $512$ & $1024$ & $1024$ \\
     Learning rate & $2e-2$ & $1e-2$ & $5e-3$ & $5e-2$ & $5e-4$ \\
     Diffusion steps & $100$ & $100$ & $100$ & $1000$ & $100$ \\
     \bottomrule
\end{tabularx}
% }
\end{table}
}

\subsection{Hyperparameters for Different Feature Selection Algorithms}
The \privmrf and \aim selection algorithms are set to completely the same as the original work in \privmrf and \aim, respectively. Therefore, we omit the description of them here and provide the detailed hyperparameter setting of \rapp selection and \privsyn selection.

{
\setlength{\abovecaptionskip}{3pt}
\begin{table}[H]
\footnotesize
\centering
\caption{\rapp Selection Hyperparameters}
% \resizebox{\columnwidth}{!}{
\begin{tabularx}{0.75\columnwidth}{l|XXXXX}
     \toprule
     \textbf{Hyperparameter} & \textbf{INC}& \textbf{EMP}& \textbf{BK}& \textbf{HIG} & \textbf{LN}\\
     \midrule
     Top q & $3$ & $3$ & $3$ & $3$ & $3$ \\
     Selection step & $4$ & $6$ & $6$ & $7$ & $10$ \\
     Selection budget rate & $0.5$ & $0.5$ & $0.5$ & $0.5$ & $0.5$ \\
     Marginal budget rate & $0.5$ & $0.5$ & $0.5$ & $0.5$ & $0.5$ \\
     \bottomrule
\end{tabularx}
% }
\end{table}
}

{
\setlength{\abovecaptionskip}{3pt}
\begin{table}[H]
\footnotesize
\centering
\caption{\privsyn Selection Hyperparameters}
% \resizebox{\columnwidth}{!}{
\begin{tabularx}{0.75\columnwidth}{l|XXXXX}
     \toprule
     \textbf{Hyperparameter} & \textbf{INC}& \textbf{EMP}& \textbf{BK}& \textbf{HIG} & \textbf{LN}\\
     \midrule
     Selection budget rate & $0.1$ & $0.1$ & $0.1$ & $0.1$ & $0.1$ \\
     1-way marginal budget rate & $0.1$ & $0.1$ & $0.1$ & $0.1$ & $0.1$ \\
     2-way marginal budget rate & $0.8$ & $0.8$ & $0.8$ & $0.8$ & $0.8$ \\
     \bottomrule
\end{tabularx}
% }
\end{table}
}

\subsection{Hyperparameters for Different Synthesis Algorithms}

Most synthesis algorithms we used are set to be the same as their original works, while relaxed projection and generative network methods need hyperparameter tuning to guarantee performance. 

{
\setlength{\abovecaptionskip}{3pt}
\begin{table}[H]
\footnotesize
\centering
\caption{Relaxed Projection Hyperparameters}
\resizebox{0.75\columnwidth}{!}{
\begin{tabular}{l|ccccc}
     \toprule
     \textbf{Hyperparameter} & \textbf{INC}& \textbf{EMP}& \textbf{BK}& \textbf{HIG} & \textbf{LN}\\
     \midrule
     Random projection number & $2e+6$ & $2e+6$ & $2e+6$ & $2e+6$ & $2e+6$ \\
     Optimization rate & $5e-3$ & $5e-3$ & $5e-3$ & $5e-3$ & $5e-3$ \\
     Optimization step & $100$ & $170$ & $160$ & $280$ & $420$ \\
     \bottomrule
\end{tabular}
}
\end{table}
}

{
\setlength{\abovecaptionskip}{3pt}
\begin{table}[H]
\footnotesize
\centering
\caption{Generative Network Hyperparameters}
\resizebox{0.75\columnwidth}{!}{
\begin{tabular}{l|ccccc}
     \toprule
     \textbf{Hyperparameter} & \textbf{INC}& \textbf{EMP}& \textbf{BK}& \textbf{HIG} & \textbf{LN}\\
     \midrule
     Learning rate & $1e-3$ & $1e-3$ & $1e-3$ & $1e-3$ & $1e-3$ \\
     Synthesis size & $1024$ & $1024$ & $1024$ & $1024$ & $1024$ \\
     Max training iteration & $50$ & $50$ & $100$ & $100$ & $1500$ \\
     \bottomrule
\end{tabular}
}
\end{table}
}

{
\setlength{\abovecaptionskip}{3pt}
\begin{table*}[t]
\footnotesize
    \centering
    \setlength{\tabcolsep}{4.5pt}
    \caption{Supplementary overall results of synthetic data under different methods. ML AUC and ML Accuracy are metrics obtained by downstream ML tasks. Running time is the total average execution time of the algorithm. Because Loan dataset is a multi-classification problem, thus it does not have AUC result.}
    \resizebox{1.0\textwidth}{!}{
    \begin{tabular}{l|ccc|ccc|ccc|ccc|ccc}
    \toprule 
        \textbf{Dataset} & \multicolumn{3}{c|}{\textbf{ACSincome}} & \multicolumn{3}{c|}{\textbf{ACSemploy}} & \multicolumn{3}{c|}{\textbf{Bank}} & \multicolumn{3}{c|}{\textbf{Higgs-small}} & \multicolumn{3}{c}{\textbf{Loan}}\\ 
        \cmidrule{1-16}
        \textbf{ML AUC} & $\varepsilon = 0.2$ & $\varepsilon = 1$ & $\varepsilon = 5$ & $\varepsilon = 0.2$ & $\varepsilon = 1$ & $\varepsilon = 5$ & $\varepsilon = 0.2$ & $\varepsilon = 1$ & $\varepsilon = 5$ & $\varepsilon = 0.2$ & $\varepsilon = 1$ & $\varepsilon = 5$ & $\varepsilon = 0.2$ & $\varepsilon = 1$ & $\varepsilon = 5$ \\ 
        \midrule
        \privsyn & 0.53 & 0.50 & 0.51 & 0.47 & 0.47 & 0.42 & 0.44 & 0.52 & 0.49 & 0.50 & 0.50 & 0.50 & - & - & - \\
        \privmrf & 0.82 & 0.87 & 0.87 & 0.80 & 0.86 & 0.89 & 0.71 & 0.90 & 0.92 & 0.53 & 0.71 & 0.70 & - & - & - \\
        \rapp & 0.75 & 0.82 & 0.85 & 0.81 & 0.84 & 0.87 & 0.75 & 0.85 & 0.88 & 0.54 & 0.56 & 0.56 & - & - & - \\
        \aim & 0.85 & 0.87 & 0.87 & 0.85 & 0.88 & 0.88 & 0.87 & 0.89 & 0.91 & 0.68 & 0.72 & 0.74 & - & - & - \\
        \gsd & 0.85 & 0.85 & 0.85 & 0.78 & 0.80 & 0.79 & 0.68 & 0.67 & 0.67 & 0.52 & 0.52 & 0.52 & - & - & - \\
        \gem & 0.77 & 0.73 & 0.72 & 0.75 & 0.77 & 0.77 & 0.59 & 0.68 & 0.69 & 0.55 & 0.57 & 0.59 & - & - & - \\
        \merf & 0.75 & 0.78 & 0.79 & 0.73 & 0.78 & 0.74 & 0.72 & 0.64 & 0.65 & 0.54 & 0.60 & 0.60 & - & - & - \\
        \ddpm & 0.54 & 0.49 & 0.53 & 0.56 & 0.56 & 0.55 & 0.48 & 0.51 & 0.45 & 0.51 & 0.53 & 0.53 & - & - & - \\
    \bottomrule
    \toprule 
        \textbf{ML Accuracy} & $\varepsilon = 0.2$ & $\varepsilon = 1$ & $\varepsilon = 5$ & $\varepsilon = 0.2$ & $\varepsilon = 1$ & $\varepsilon = 5$ & $\varepsilon = 0.2$ & $\varepsilon = 1$ & $\varepsilon = 5$ & $\varepsilon = 0.2$ & $\varepsilon = 1$ & $\varepsilon = 5$ & $\varepsilon = 0.2$ & $\varepsilon = 1$ & $\varepsilon = 5$ \\ 
        \midrule
        \privsyn & 0.59 & 0.58 & 0.59 & 0.52 & 0.49 & 0.49 & 0.88 & 0.88 & 0.88 & 0.53 & 0.52 & 0.52 & 0.54 & 0.54 & 0.54 \\
        \privmrf & 0.75 & 0.79 & 0.79 & 0.72 & 0.79 & 0.81 & 0.89 & 0.90 & 0.90 & 0.53 & 0.65 & 0.64 & 0.76 & 0.75 & 0.76 \\
        \rapp & 0.68 & 0.75 & 0.78 & 0.74 & 0.78 & 0.80 & 0.85 & 0.88 & 0.89 & 0.53 & 0.55 & 0.55 & 0.62 & 0.64 & 0.65 \\
        \aim & 0.78 & 0.79 & 0.79 & 0.78 & 0.80 & 0.81 & 0.90 & 0.90 & 0.90 & 0.63 & 0.65 & 0.67 & 0.75 & 0.75 & 0.75 \\
        \gsd & 0.77 & 0.77 & 0.78 & 0.71 & 0.73 & 0.71 & 0.87 & 0.87 & 0.88 & 0.51 & 0.51 & 0.52 & 0.54 & 0.54 & 0.54 \\
        \gem & 0.71 & 0.69 & 0.66 & 0.68 & 0.70 & 0.69 & 0.86 & 0.86 & 0.87 & 0.55 & 0.55 & 0.56 & 0.71 & 0.68 & 0.72 \\
        \merf & 0.69 & 0.70 & 0.72 & 0.64 & 0.71 & 0.67 & 0.84 & 0.79 & 0.74 & 0.53 & 0.58 & 0.58 & 0.35 & 0.37 & 0.37 \\
        \ddpm & 0.57 & 0.59 & 0.59 & 0.55 & 0.54 & 0.53 & 0.88 & 0.88 & 0.88 & 0.50 & 0.53 & 0.51 & 0.55 & 0.55 & 0.55 \\
    \bottomrule
    \toprule
        \textbf{Running Time (min)} & $\varepsilon = 0.2$ & $\varepsilon = 1$ & $\varepsilon = 5$ & $\varepsilon = 0.2$ & $\varepsilon = 1$ & $\varepsilon = 5$ & $\varepsilon = 0.2$ & $\varepsilon = 1$ & $\varepsilon = 5$ & $\varepsilon = 0.2$ & $\varepsilon = 1$ & $\varepsilon = 5$ & $\varepsilon = 0.2$ & $\varepsilon = 1$ & $\varepsilon = 5$ \\ 
        \midrule
        \privsyn & 0.28 & 0.16 & 0.18 & 0.30 & 0.25 & 0.34 & 0.57 & 0.53 & 0.55 & 5.39 & 5.95 & 4.54 & 13.29 & 14.04 & 13.44 \\
        \privmrf & 1.26 & 0.89 & 1.40 & 5.59 & 5.80 & 4.76 & 3.06 & 4.48 & 3.24 & 5.06 & 7.10 & 6.25 & 7.12 & 13.34 & 12.62 \\
        \rapp & 25.86 & 24.95 & 25.72 & 26.28 & 27.04 & 28.30 & 24.40 & 23.94 & 23.27 & 37.79 & 36.66 & 35.88 & 544.25 & 2348.59 & 1761.60 \\
        \aim & 1.96 & 6.29 & 126.89 & 5.37 & 10.59 & 443.70 & 3.60 & 10.23 & 23.57 & 12.62 & 18.46 & 184.57 & 31.03 & 187.27 & 645.13 \\
        \gsd & 22.33 & 24.86 & 26.17 & 10.89 & 11.12 & 11.11 & 19.38 & 20.01 & 19.99 & 54.35 & 57.49 & 57.93 & 304.42 & 306.04 & 311.63 \\
        \gem & 0.26 & 0.10 & 0.09 & 0.12 & 0.12 & 0.12 & 0.27 & 0.25 & 0.25 & 6.05 & 6.09 & 10.75 & 9.82 & 9.47 & 9.82 \\
        \merf & 0.40 & 0.07 & 0.06 & 0.30 & 0.06 & 0.06 & 0.09 & 0.06 & 0.06 & 0.09 & 0.06 & 0.06 & 0.39 & 0.83 & 0.34 \\
        \ddpm & 4.34 & 4.12 & 3.73 & 4.52 & 4.50 & 13.31 & 8.25 & 4.40 & 3.87 & 4.74 & 6.34 & 4.78 & 27.54 & 31.93 & 33.34 \\
    \bottomrule
    \toprule
        \textbf{Conditional Query Error} & $\varepsilon = 0.2$ & $\varepsilon = 1$ & $\varepsilon = 5$ & $\varepsilon = 0.2$ & $\varepsilon = 1$ & $\varepsilon = 5$ & $\varepsilon = 0.2$ & $\varepsilon = 1$ & $\varepsilon = 5$ & $\varepsilon = 0.2$ & $\varepsilon = 1$ & $\varepsilon = 5$ & $\varepsilon = 0.2$ & $\varepsilon = 1$ & $\varepsilon = 5$ \\ 
        \midrule
        \privsyn & 0.063 & 0.019 & 0.017 & 0.048 & 0.035 & 0.030 & 0.034 & 0.033 & 0.030 & 0.034 & 0.029 & 0.024 & 0.070 & 0.058 & 0.052 \\
        \privmrf & 0.014 & 0.012 & 0.009 & 0.022 & 0.013 & 0.010 & 0.031 & 0.026 & 0.024 & 0.025 & 0.026 & 0.025 & 0.054 & 0.053 & 0.051 \\
        \rapp & 0.029 & 0.021 & 0.020 & 0.039 & 0.026 & 0.019 & 0.072 & 0.064 & 0.063 & 0.099 & 0.098 & 0.104 & 0.108 & 0.098 & 0.092 \\
        \aim & 0.030 & 0.013 & 0.010 & 0.028 & 0.011 & 0.008 & 0.033 & 0.022 & 0.022 & 0.028 & 0.025 & 0.025 & 0.058 & 0.057 & 0.050 \\
        \gsd & 0.024 & 0.016 & 0.011 & 0.027 & 0.014 & 0.009 & 0.045 & 0.039 & 0.037 & 0.110 & 0.110 & 0.107 & 0.091 & 0.093 & 0.096 \\
        \gem & 0.025 & 0.026 & 0.026 & 0.042 & 0.028 & 0.026 & 0.055 & 0.053 & 0.051 & 0.068 & 0.069 & 0.073 & 0.077 & 0.077 & 0.077 \\
        \merf & 0.308 & 0.308 & 0.311 & 0.316 & 0.314 & 0.313 & 0.210 & 0.215 & 0.238 & 0.101 & 0.099 & 0.104 & 0.416 & 0.455 & 0.426 \\
        \ddpm & 0.215 & 0.092 & 0.083 & 0.268 & 0.072 & 0.079 & 0.131 & 0.136 & 0.144 & 0.566 & 0.109 & 0.109 & 0.143 & 0.143 & 0.140 \\
    \bottomrule
    \end{tabular}
    }
    \label{extra general result}
\end{table*}
}

\section{Supplementary Experiment Results}
\label{appendix: res}

\subsection{More Results of Algorithm Comparison}

We have provided some important metric results of algorithms' utility in \cref{overall exp}. Here we list some other detailed evaluations, such as AUC, accuracy for machine learning efficiency, and running time for time efficiency. Notably, these results do not necessarily influence our conclusion. Thus, we present them in the appendix.

\subsection{Detailed Results of Preprocessing Influence}
The detailed results of different preprocessing methods (used to plot \cref{fig2: exp}) are shown in \cref{prep detail result}. Similar to other experiments, these metrics are calculated by comparing preprocessed datasets with test datasets to demonstrate the error caused by preprocessing.

{
\setlength{\abovecaptionskip}{3pt}
\begin{table}[t]
\footnotesize
\centering
\caption{\Rtwo{Comparison of fidelity error and Cramer's V measure error on Bank dataset under $\varepsilon = 0.2$.}}
\label{cramer}
\resizebox{0.7\columnwidth}{!}{
\begin{tabular}{l|cc|l|cc}
    \toprule
    \textbf{Method} & \textbf{Fid Err.} & \textbf{Cramer Err.} & \textbf{Method} & \textbf{Fid Err.} & \textbf{Cramer Err.} \\ 
    \midrule
    Privsyn & 0.24 & 0.10 & Private-GSD & 0.52 & 0.11 \\
    PrivMRF & 0.13 & 0.07 & GEM & 0.24 & 0.10 \\
    RAP++ & 0.43 & 0.10 & DP-MERF & 0.47 & 0.11 \\
    AIM & 0.11 & 0.06 & TabDDPM & 0.77 & 0.13 \\
    \bottomrule
\end{tabular}
}
\end{table}
}

{
\setlength{\abovecaptionskip}{3pt}
\begin{table*}[!h]
\footnotesize
    \centering
    \caption{Influences of preprocessing on different datasets. By default, the results are obtained under the setting that $\varepsilon = 1.0$ and $10\%$ of the budget is used for preprocessing.}
    \label{prep detail result}
    \setlength{\tabcolsep}{4.5pt}
    \resizebox{1.0\textwidth}{!}{
    \begin{tabular}{l|ccc|ccc|ccc|ccc}
    \toprule 
        \multirow{2}{*}{\textbf{Preprocessing Method}} & \multicolumn{3}{c|}{\textbf{Marginal Size}} & \multicolumn{3}{c|}{\textbf{ML Efficacy}} & \multicolumn{3}{c|}{\textbf{Query Error}} & \multicolumn{3}{c}{\textbf{Fidelity Error}}\\ \cline{2-13}
        & \rule{0pt}{1.1em} Bank & Higgs-small & Loan & Bank & Higgs-small & Loan & Bank & Higgs-small & Loan & Bank & Higgs-small & Loan \\ 
        \midrule
        Raw                       & $1.05e+5$& $4.24e+8$& $2.88e+8$& $0.76$& $0.72$& $0.54$& $0.001$& $0.001$& $0.001$ & $0.003$ & $0.001$& $0.001$\\
        Numerical preprocessing   & $6.48e+2$& $5.05e+3$& $1.14e+6$& $0.76$& $0.71$& $0.53$& $0.002$& $0.002$& $0.003$ & $0.006$& $0.002$& $0.003$\\
        Categorical Preprocessing & $1.05e+5$& $4.24e+8$& $2.31e+8$& $0.76$& $0.72$& $0.54$& $0.001$& $0.001$& $0.001$ & $0.003$& $0.001$& $0.001$\\
        Full Preprocessing        & $6.48e+2$& $5.05e+3$& $3.14e+3$& $0.76$& $0.71$& $0.53$& $0.002$& $0.002$& $0.003$ & $0.006$& $0.002$& $0.003$\\
    \bottomrule
    \end{tabular}
    }
\end{table*}
}

{
\setlength{\abovecaptionskip}{3pt}
\begin{table*}[!h]
\footnotesize
    \centering
    \caption{Results of synthetic data under different feature selection methods. By default, we use \mtdtt{PGM} as the synthesis method. In this table, ``-'' means unable to execute due to time or memory limitation.}
    \label{select detail result}
    \setlength{\tabcolsep}{4.5pt}
    \resizebox{1.0\textwidth}{!}{
    \begin{tabular}{l|ccc|ccc|ccc|ccc|ccc}
    \toprule 
        \textbf{Dataset} & \multicolumn{3}{c|}{\textbf{ACSincome}} & \multicolumn{3}{c|}{\textbf{ACSemploy}} & \multicolumn{3}{c|}{\textbf{Bank}} & \multicolumn{3}{c|}{\textbf{Higgs-small}} & \multicolumn{3}{c}{\textbf{Loan}}\\ \midrule
        \textbf{ML Efficacy} & $\varepsilon = 0.2$ & $\varepsilon = 1$ & $\varepsilon = 5$ & $\varepsilon = 0.2$ & $\varepsilon = 1$ & $\varepsilon = 5$ & $\varepsilon = 0.2$ & $\varepsilon = 1$ & $\varepsilon = 5$ & $\varepsilon = 0.2$ & $\varepsilon = 1$ & $\varepsilon = 5$ & $\varepsilon = 0.2$ & $\varepsilon = 1$ & $\varepsilon = 5$ \\ 
        \midrule
        \privsyn selection & $0.67$ & $0.65$ & $0.68$ & $0.67$ & $0.80$ & $0.75$ & $-$ & $-$ & $-$ & $-$ & $-$ & $-$ & $-$ & $-$ & $-$\\
        \privmrf selection & $0.73$ & $0.78$ & $0.78$ & $0.81$ & $0.80$ & $0.81$ & $0.62$ & $0.70$ & $0.71$ & $0.50$ & $0.64$ & $0.64$ & $0.52$ & $0.52$ & $0.52$ \\
        \rapp selection     & $0.62$ & $0.74$ & $0.74$ & $0.74$ & $0.78$ & $0.79$ & $0.51$ & $0.51$ & $0.47$ & $0.49$ & $0.49$ & $0.50$ & $0.30$ & $0.26$ & $0.26$ \\
        \aim selection      & $0.76$ & $0.78$ & $0.78$ & $0.78$ & $0.80$ & $0.81$ & $0.67$ & $0.71$ & $0.70$ & $0.63$ & $0.65$ & $0.67$ & $0.52$ & $0.52$ & $0.52$ \\
    \bottomrule
    \toprule 
        \textbf{Query Error} & $\varepsilon = 0.2$ & $\varepsilon = 1$ & $\varepsilon = 5$ & $\varepsilon = 0.2$ & $\varepsilon = 1$ & $\varepsilon = 5$ & $\varepsilon = 0.2$ & $\varepsilon = 1$ & $\varepsilon = 5$ & $\varepsilon = 0.2$ & $\varepsilon = 1$ & $\varepsilon = 5$ & $\varepsilon = 0.2$ & $\varepsilon = 1$ & $\varepsilon = 5$ \\ 
        \midrule
        \privsyn selection & $0.003$ & $0.001$ & $0.001$ & $0.003$ & $0.002$ & $0.003$ & $-$ & $-$ & $-$ & $-$ & $-$ & $-$ & $-$ & $-$ & $-$\\
        \privmrf selection & $0.002$ & $0.001$ & $0.001$ & $0.003$ & $0.002$ & $0.002$ & $0.005$ & $0.003$ & $0.003$ & $0.005$ & $0.003$ & $0.003$ & $0.005$ & $0.005$ & $0.004$ \\
        \rapp selection     & $0.006$ & $0.003$ & $0.002$ & $0.008$ & $0.005$ & $0.004$ & $0.012$ & $0.005$ & $0.003$ & $0.047$ & $0.016$ & $0.006$ & $0.015$ & $0.009$ & $0.008$ \\
        \aim selection      & $0.002$ & $0.001$ & $0.001$ & $0.003$ & $0.002$ & $0.001$ & $0.007$ & $0.002$ & $0.002$ & $0.005$ & $0.003$ & $0.003$ & $0.005$ & $0.005$ & $0.004$ \\
    \bottomrule
    \toprule 
        \textbf{Fidelity Error} & $\varepsilon = 0.2$ & $\varepsilon = 1$ & $\varepsilon = 5$ & $\varepsilon = 0.2$ & $\varepsilon = 1$ & $\varepsilon = 5$ & $\varepsilon = 0.2$ & $\varepsilon = 1$ & $\varepsilon = 5$ & $\varepsilon = 0.2$ & $\varepsilon = 1$ & $\varepsilon = 5$ & $\varepsilon = 0.2$ & $\varepsilon = 1$ & $\varepsilon = 5$ \\ 
        \midrule
        \privsyn selection  & $0.14$ & $0.08$ & $0.07$ & $0.07$ & $0.04$ & $0.04$ & $-$ & $-$ & $-$ & $-$ & $-$ & $-$ & $-$ & $-$ & $-$\\
        \privmrf selection  & $0.11$ & $0.07$ & $0.05$ & $0.07$ & $0.04$ & $0.03$ & $0.13$ & $0.06$ & $0.04$ & $0.36$ & $0.19$ & $0.19$ & $0.31$ & $0.24$ & $0.23$ \\
        \rapp selection     & $0.17$ & $0.09$ & $0.08$ & $0.04$ & $0.04$ & $0.04$ & $0.14$ & $0.09$ & $0.08$ & $0.42$ & $0.23$ & $0.17$ & $0.32$ & $0.26$ & $0.25$ \\
        \aim selection      & $0.09$ & $0.06$ & $0.05$ & $0.05$ & $0.03$ & $0.02$ & $0.11$ & $0.09$ & $0.06$ & $0.21$ & $0.16$ & $0.16$ & $0.35$ & $0.32$ & $0.29$\\
    \bottomrule
    \end{tabular}
    }
\end{table*}
}

{
\setlength{\abovecaptionskip}{3pt}
\begin{table*}[t]
\footnotesize
    \centering
    \caption{Results of synthetic data under different synthesis methods. By default, we use \privsyn's InDif selection as the selection method. In this table, ``-'' means unable to execute due to time or memory limitation.}
    \label{syn detail result}
    \setlength{\tabcolsep}{4.5pt}
    \resizebox{1.0\textwidth}{!}{
    \begin{tabular}{l|ccc|ccc|ccc|ccc|ccc}
    \toprule 
        \textbf{Dataset} & \multicolumn{3}{c|}{\textbf{ACSincome}} & \multicolumn{3}{c|}{\textbf{ACSemploy}} & \multicolumn{3}{c|}{\textbf{Bank}} & \multicolumn{3}{c|}{\textbf{Higgs-small}} & \multicolumn{3}{c}{\textbf{Loan}}\\ 
    \midrule
        \textbf{ML Efficacy} & $\varepsilon = 0.2$ & $\varepsilon = 1$ & $\varepsilon = 5$ & $\varepsilon = 0.2$ & $\varepsilon = 1$ & $\varepsilon = 5$ & $\varepsilon = 0.2$ & $\varepsilon = 1$ & $\varepsilon = 5$ & $\varepsilon = 0.2$ & $\varepsilon = 1$ & $\varepsilon = 5$ & $\varepsilon = 0.2$ & $\varepsilon = 1$ & $\varepsilon = 5$ \\ 
    \midrule
        \mtdtt{GUM}                   & $0.39$ & $0.40$ & $0.42$ & $0.45$ & $0.45$ & $0.40$ & $0.47$ & $0.47$ & $0.47$ & $0.40$ & $0.43$ & $0.43$ & $0.25$ & $0.26$ & $0.26$ \\
        \mtdtt{PGM}                   & $0.76$ & $0.64$ & $0.67$ & $0.68$ & $0.80$ & $0.75$ & $-$ & $-$ & $-$ & $-$ & $-$ & $-$ & $-$ & $-$ & $-$ \\
        Relaxed Projection           & $0.39$ & $0.59$ & $0.58$ & $0.51$ & $0.69$ & $0.67$ & $0.47$ & $0.47$ & $0.47$ & $0.47$ & $0.44$ & $0.42$ & $0.25$ & $0.25$ & $0.25$  \\
        Genetic Algorithm            & $0.67$ & $0.63$ & $0.58$ & $0.62$ & $0.58$ & $0.62$ & $0.61$ & $0.57$ & $0.52$ & $0.59$ & $0.58$ & $0.62$ & $0.35$ & $0.36$ & $0.26$  \\
        Generative Network           & $0.74$ & $0.77$ & $0.76$ & $0.72$ & $0.70$ & $0.69$ & $0.66$ & $0.63$ & $0.62$ & $0.63$ & $0.53$ & $0.59$ & $0.46$ & $0.41$ & $0.36$  \\
    \bottomrule
    \toprule 
        \textbf{Query Error} & $\varepsilon = 0.2$ & $\varepsilon = 1$ & $\varepsilon = 5$ & $\varepsilon = 0.2$ & $\varepsilon = 1$ & $\varepsilon = 5$ & $\varepsilon = 0.2$ & $\varepsilon = 1$ & $\varepsilon = 5$ & $\varepsilon = 0.2$ & $\varepsilon = 1$ & $\varepsilon = 5$ & $\varepsilon = 0.2$ & $\varepsilon = 1$ & $\varepsilon = 5$ \\ 
    \midrule
        \mtdtt{GUM}                   & $0.003$ & $0.002$ & $0.002$ & $0.006$ & $0.004$ & $0.004$ & $0.007$ & $0.004$ & $0.003$ & $0.009$ & $0.004$ & $0.003$ & $0.006$ & $0.005$ & $0.004$ \\
        \mtdtt{PGM}                   & $0.003$ & $0.001$ & $0.001$ & $0.003$ & $0.002$ & $0.003$ & $-$ & $-$ & $-$ & $-$ & $-$ & $-$ & $-$ & $-$ & $-$  \\
        Relaxed Projection           & $0.040$ & $0.028$ & $0.023$ & $0.081$ & $0.025$ & $0.022$ & $0.047$ & $0.013$ & $0.014$ & $0.026$ & $0.013$ & $0.012$ & $0.020$ & $0.009$ & $0.008$  \\
        Genetic Algorithm            & $0.053$ & $0.047$ & $0.050$ & $0.039$ & $0.040$ & $0.036$ & $0.009$ & $0.005$ & $0.003$ & $0.027$ & $0.018$ & $0.016$ & $0.039$ & $0.037$ & $0.039$  \\
        Generative Network           & $0.002$ & $0.002$ & $0.002$ & $0.003$ & $0.003$ & $0.003$ & $0.003$ & $0.003$ & $0.003$ & $0.003$ & $0.003$ & $0.003$ & $0.003$ & $0.004$ & $0.004$  \\
    \bottomrule
    \toprule 
        \textbf{Fidelity Error} & $\varepsilon = 0.2$ & $\varepsilon = 1$ & $\varepsilon = 5$ & $\varepsilon = 0.2$ & $\varepsilon = 1$ & $\varepsilon = 5$ & $\varepsilon = 0.2$ & $\varepsilon = 1$ & $\varepsilon = 5$ & $\varepsilon = 0.2$ & $\varepsilon = 1$ & $\varepsilon = 5$ & $\varepsilon = 0.2$ & $\varepsilon = 1$ & $\varepsilon = 5$ \\ 
    \midrule
        \mtdtt{GUM}                   & $0.15$ & $0.12$ & $0.12$ & $0.12$ & $0.09$ & $0.09$ & $0.24$ & $0.13$ & $0.12$ & $0.29$ & $0.20$ & $0.20$ & $0.34$ & $0.35$ & $0.36$ \\
        \mtdtt{PGM}                   & $0.13$ & $0.08$ & $0.07$ & $0.07$ & $0.04$ & $0.04$ & $-$ & $-$ & $-$ & $-$ & $-$ & $-$ & $-$ & $-$ & $-$ \\
        Relaxed Projection           & $0.61$ & $0.52$ & $0.41$ & $0.51$ & $0.26$ & $0.21$ & $0.41$ & $0.16$ & $0.16$ & $0.31$ & $0.22$ & $0.21$ & $0.41$ & $0.29$ & $0.28$  \\
        Genetic Algorithm            & $0.59$ & $0.59$ & $0.56$ & $0.30$ & $0.30$ & $0.30$ & $0.13$ & $0.09$ & $0.08$ & $0.35$ & $0.28$ & $0.27$ & $0.56$ & $0.55$ & $0.55$  \\
        Generative Network           & $0.08$ & $0.08$ & $0.08$ & $0.06$ & $0.07$ & $0.06$ & $0.16$ & $0.16$ & $0.15$ & $0.24$ & $0.23$ & $0.23$ & $0.47$ & $0.38$ & $0.27$  \\
    \bottomrule
    \end{tabular}
    }
\end{table*}
}

\subsection{Detailed Results of Reconstruction Experiment}
The detailed results of different preprocessing methods (used to plot \cref{fig3: exp}) are shown in \cref{select detail result} and \cref{syn detail result}. Here, there are some ``-'' in the table. This is because PrivSyn tends to select as many marginals as possible, which will form large cliques. This will cause the size of the graphical model to be too large, requiring extremely large memory.

\subsection{Discussion for Cramer's V measure}

\Rtwo{We further explain why we do not include Cramer's V measure in our evaluation metrics with some experimental results. In \Cref{cramer}, a key finding is that even though some methods vary largely in fidelity error, their evaluations of Cramer's V measure error can be close. For example, \gem and \rapp both have a result of 0.1 on Cramer's V measure error, but their fidelity errors actually differ. This result supports our argument in \Cref{sec: exp setup}.}

\end{document}